\documentclass[sn-aps]{sn-jnl}

\usepackage{etoolbox}
\makeatletter
\patchcmd{\ps@headings}
{\hbox to \hsize{\hfill Springer Nature 2021 \LaTeX\ template\hfill}}
{\hbox to \hsize{}}
{}
{}
\patchcmd{\ps@headings}
{\hbox to \hsize{\hfill Springer Nature 2021 \LaTeX\ template\hfill}}
{\hbox to \hsize{}}
{}
{}
\patchcmd{\ps@titlepage}
{\hbox to \hsize{\hfill Springer Nature 2021 \LaTeX\ template\hfill}}
{\hbox to \hsize{}}
{}
{}
\makeatother

\jyear{2021}%

\begin{document}

\title[RMPP]{\bf Self-organization of photoionized plasmas via kinetic instabilities}

\author*[1]{\fnm{Chaojie} \sur{Zhang}}\email{chaojiez@ucla.edu}

\author[1,2]{\fnm{Chen-Kang} \sur{Huang}}\email{ckhuang@ucla.edu}

\author*[1]{\fnm{Chan} \sur{Joshi}}\email{cjoshi@ucla.edu}

\affil[1]{\orgdiv{Department of Electrical and Computer Engineering}, \orgname{UCLA}, \orgaddress{\city{Los Angeles}, \postcode{90095}, \state{CA}, \country{USA}}}
\affil[2]{\orgdiv{Institute of Atomic and Molecular Sciences}, \orgname{Academia Sinica}, \orgaddress{\city{Taipei}, \postcode{10617}, \country{Taiwan}}}

\abstract{Self-organization in an unmagnetized collisionless plasma (in this paper) refers to formation of transient coherent structures such as collective oscillations (electrostatic waves) or magnetic fields resulting from so-called kinetic effects in the plasma. This topical review provides a comprehensive analysis of the self-organization of strong-field photoionized, non-equilibrium plasmas through kinetic instabilities. The authors propose and demonstrate a novel experimental platform that enables the formation of dense plasmas with known highly anisotropic and non-thermal electron velocity distribution functions on a timescale on the order of an inverse electron plasma frequency. We then show that such plasmas are highly susceptible to a hierarchy of kinetic instabilities, including two-stream, current filamentation and Weibel, that convert a fraction of the electron kinetic energy into electric and/or magnetic energy stored in self-organized structures. The electrostatic waves so produced are measured using a collective light (Thomson) scattering technique with femtosecond resolution as the kinetic instabilities aided by collisions eventually thermalize the plasma electrons. In addition, we describe a novel experimental technique that has made it possible to map the temporal evolution of the wavenumber spectrum of the thermal Weibel instability with picosecond resolution, which leads to the formation of quasistatic coherent magnetic fields with different topologies in photoionized plasmas. Finally, the paper summarizes the important results and discusses future directions on this topic.}




\maketitle

\section{Introduction}\label{sec1}
Self-organization in a system usually refers to the spontaneous appearance of structures with increasing complexity as a result of one or more sequential or concurrent instabilities \cite{turing_a_chemical_1952,nicolis_self-organization_1977,klimontovich_statistical_1994}. In plasmas, self-organization can spontaneously occur and spread through the interaction of charged particles with collective fields- often called wave-particle interactions. An early example of spontaneous growth of a plasma wave was the theoretical prediction \cite{gould_plasma_1967,malmberg_plasma_1968} and experimental demonstration of the plasma-wave echo phenomenon enabled by wave-particle interactions henceforth called kinetic effects. This mechanism, an example of inverse Landau damping arising from a non-thermal electron velocity (often called bump-on-tail distribution \cite{tsunoda_experimental_1991}), is perhaps the best-known example of a kinetic instability that can spontaneously cause growth of a plasma wave through self-organization.

The field of kinetic plasma instabilities \cite{chen_introduction_2016,pierce_possible_1948,bohm_theory_1949,bohm_theory_1949-2,weibel_spontaneously_1959,fried_mechanism_1959} is arguably as old as our attempts to describe plasma phenomena itself as summarized in Fig. 1. A fundamental feature that distinguishes plasmas from ordinary matter, such as gases, fluids and solids, is the existence of collective interaction between charged particles (self-fields) under the influence of long-range electromagnetic forces (external fields). How plasmas evolve under self- and/or external applied forces is essential to accurate prediction of plasma behavior.

Perhaps the most intuitive and accurate way of describing a plasma is to follow the trajectory of individual particles in the presence of external and self-fields. The challenge, however, is that a plasma almost always involves numerous particles therefore making this approach impractical. Alternatively, a plasma can be described statistically through distribution functions, which give the number of particles per unit volume in phase space (e.g., position-velocity). This approach is the basis of plasma kinetic theory. For instance, the evolution of a collisionless plasma can be described by following its distribution function governed by the Vlasov equation. The next level of simplification is the fluid theory, which is achieved by taking the moments of the Vlasov equation over velocity space to get a set of differential equations of macroscopic quantities such as density, velocity, and temperature that govern the conservation of mass, momentum, and energy. Since the fluid theory is less complex, it does not capture all the physics, particularly when the plasma has complicated distribution functions. Kinetic theory on the other hand offers a more detailed description of the plasma behavior but at a cost of increased computational complexity.

\begin{figure}[h]%
\centering
\includegraphics[width=0.95\textwidth]{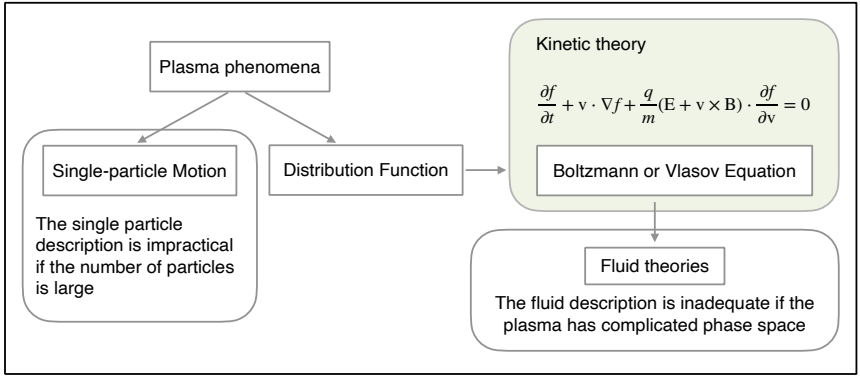}
\caption{Hierarchy description of plasma phenomena.}
\label{fig-hierarchy}
\end{figure}

The study of kinetic plasma instabilities is a major aspect of kinetic theory. Deviations of velocity distribution functions of plasma electrons, ions, or both from thermal distributions can result in the growth of kinetic instabilities \cite{chen_introduction_2016}. For instance, Bohm and Gross developed a kinetic theory of unstable perturbations propagating along the beam direction in a beam-plasma system, first suggested by Langmuir and subsequently observed by Pierce, which is now known as the “two-stream instability (TSI)” \cite{pierce_possible_1948,bohm_theory_1949,bohm_theory_1949-2}. In 1959, Weibel demonstrated that in a stationary plasma with a temperature anisotropy (i.e., temperatures differ in different spatial directions), magnetic fields can be self-generated \cite{weibel_spontaneously_1959}. This was later named as “Weibel instability (WI)”. In the same year, Fried explained the mechanism of WI- using a configuration of two identical counter-propagating beams- which showed that such a system can also become unstable against electromagnetic modulations normal to the beam direction, leading to the formation of current filaments on kinetic scales \cite{fried_mechanism_1959}. This process is now known as “current filamentation instability (CFI)”, which is closely related to the thermal Weibel instability.

The TSI is electrostatic, meaning that the wavenumber of the associated waves (charge density oscillations) is parallel to the electric field (${\bf k\times E}=0$). These waves are produced by inverse Landau damping, which converts the directional kinetic energy of the plasma streams into longitudinal waves that generally have a non-zero oscillating frequency, except for two identical counter-propagating streams. The CFI may grow concurrently with the TSI. As the CFI grows, attraction (repulsion) between co- (counter-) propagating microscopic plasma currents causes them to coalesce, leading to amplification of magnetic fields surrounding the currents. In general, the CFI is not purely transverse (i.e., ${\bf k\cdot E}\neq 0$), unless the counter-propagating streams are symmetric such that they pinch at the same rate \cite{bret_how_2007,tzoufras_space-charge_2006}. Weibel instability grows via a similar mechanism as the CFI but is transverse (${\bf k\cdot E}=0$). In a beam-plasma system, in addition to the longitudinal (TSI) and the transverse (CFI/WI) modes, oblique modes with wavevectors pointing at arbitrary angles with respect to the beam propagation direction can also grow, indicating that the unstable spectrum is truly multi-dimensional \cite{bret_multidimensional_2010}.

These kinetic instabilities have been predicted and to a lesser degree detected in a variety of laboratory beam-plasma systems and astrophysical plasmas. For example, they have been studied in the context of hot-electron transport in fast-ignition inertial confinement fusion \cite{mendonca_beam_2005,sentoku_anomalous_2003}, electron cloud effects in storage rings \cite{ohmi_head-tail_2000} and proton synchrotrons \cite{cappi_electron_2002}, quark-gluon plasmas produced in heavy-ion beam collisions \cite{mrowczynski_stream_1988,arnold_apparent_2005}, collisionless shock formation driven by energetic laser-solid interaction \cite{fox_filamentation_2013,huntington_observation_2015,fiuza_electron_2020}, and plasma-based particle acceleration \cite{su_stability_1987,yan_self-induced_2009}. Beyond the laboratory, kinetic instabilities are thought to play important roles in phenomena such as the self-generation of magnetic fields in plasmas \cite{medvedev_generation_1999,nishikawa_particle_2003}, the solar corona and interplanetary medium \cite{che_how_2017}, the trapping of solar wind in the ionosphere by Earth's magnetic field \cite{marsch_kinetic_2006}, gamma ray bursts \cite{lyubarsky_are_2006}, electron-positron plasmas \cite{yang_evolution_1994,fonseca_three-dimensional_2002}, and neutrino-plasma interactions in supernova explosions \cite{silva_exact_2006}. Understanding magnetic self-organization in plasmas is critical to understanding the cosmic magnetic structures on a variety of scales and topologies in astrophysical, solar, and space plasmas \cite{durrer_cosmological_2013}.

Given the broad range of situations that give rise to kinetic instabilities a voluminous theoretical literature exists, yet few laboratory verification has been achieved due to a lack of a suitable experimental platform for initializing (defined as time $t=0$) known non-thermal/anisotropic electron velocity distributions that act as the source of kinetic energy. In the intense laser-produced plasma platform demonstrated here, these instabilities thermalize the plasma in tens of electron plasma periods requiring ultrafast probing of the associated electrostatic and electromagnetic fields. In the past, various related approaches have been developed, including laser-driven colliding plasmas, where counterpropagating ablation plasmas are generated by evaporating a solid surface by energetic laser pulses, triggering ion CFI since the energy in these flows is predominantly carried by flowing ions \cite{fox_filamentation_2013}. Proton radiography can then capture the spatiotemporal evolution of the electric and magnetic fields with typical resolutions of tens of microns and ps \cite{huntington_observation_2015,fiuza_electron_2020,ruyer_growth_2020}. This approach (and its variations \cite{swadling_measurement_2020}) has been effective in probing high-energy density plasmas, but requires access to large facilities and is limited by very low repetition rates (typically only a few shots per day). Alternatively, passing electron beams through stationary plasmas can trigger the growth of TSI and/or CFI. The electron beams can be generated from laser-solid interaction \cite{mondal_direct_2012}, laser wakefield accelerators \cite{huntington_current_2011} or linear accelerators\cite{allen_experimental_2012,san_miguel_claveria_spatiotemporal_2022}. The growth of CFI leads to the breakup of the electron beam, which can be diagnosed by measuring the transition radiation from the filamented beam as it traverses through a metallic foil \cite{allen_experimental_2012}, the magnetic fields surrounding the filaments with optical polarimetry \cite{mondal_direct_2012}, or the divergence change of the beam itself \cite{raj_probing_2020}. However, this approach generally does not give any information about the temporal growth of the instability.

The growth of TSI leads to thermalization of the electron distribution function, while the growth of CFI and WI leads to isotropization of the plasma. The latter instabilities can generate and amplify coherent electromagnetic structures in an irreversible process as the plasma electrons approaches thermal equilibrium state by collisions. The theoretical framework of kinetic instabilities shows that source of energy needed to produce self-organized structures in plasmas is the deviation of the electron velocity distribution function from an isotropic, thermal (Maxwell-Boltzmann) case.

Here, we focus on our recent experimental advancements on this topic enabled by a novel platform developed by the authors \cite{zhang_ultrafast_2019,zhang_measurements_2020,zhang_mapping_2022}. This relatively simple and extremely reproducible platform has allowed us to prepare plasmas with known initial non-thermal and highly anisotropic distribution functions and to measure density or current fluctuations using external optical or electron probes on ultrafast time scales that are otherwise inaccessible. We present experimental measurements of TSI and CFI using femtosecond Thomson scattering and of WI using ultrafast relativistic electron probing, which enabled us to validate important predictions of kinetic theory. Using this unique platform, we have also visualized the self-organization process that leads to large-scale ($\gg c/\omega_p$) and long lived ($\gg\omega_p^{-1}$) magnetic fields in photoionized plasmas due to the growth of electron thermal Weibel instability. For plasma densities of $>10^{18}~\rm cm^{-3}$ easily produced via tunnel ionization within a few fs, the above spatial and temporal constraints require diagnostic techniques that have a spatial resolution of a few microns and a temporal resolution in the 100 fs to ps range to resolve the wavenumber spectrum and linear growth rates of the kinetic processes that lead to the formation of the self-organized structures in the plasma respectively.

\section{Electron distribution functions of photoionized plasmas}\label{sec2}
In order to excite kinetic instabilities in experiments in a controlled manner, it is necessary to initialize (at time $t=0$) a plasma with a nonthermal and/or anisotropic velocity distribution(s). Here we consider two limits of generating such a plasma using photoionization. In the first case the photon energy exceeds the ionization potential. The second limit is when the photon energy is far less than the ionization potential where ionization nevertheless can happen in the so called strong-field tunneling regime. In the theory section we consider both limits but our experiments are done in the second, tunnel ionization regime.

\subsection{Single-photon photoionization}\label{sec-single-photon}
We consider here the ionization of both helium electrons by absorption of a single photon. The ionization potentials to form $\rm He^{1+}$ and $\rm He^{2+}$ ions are $I_1=24.6$ eV and $I_2=54.4$ eV, respectively. Clearly, an intense EUV free electron laser such as FLASH or Swiss FEL \cite{ackermann_operation_2007,milne_swissfel_2017,mcneil_x-ray_2010} is required to reach this regime. There is no intensity threshold for this process; it only requires the photons’ energy to exceed the ionization threshold of the second electron. The ionization rate is however proportional to the photon beam energy density. 

The overall photoionization cross section is related to the optical oscillator strength density ${\rm d}f/{\rm d}E$ by the following relation \cite{fano_spectral_1968}
\begin{equation}
\sigma=\frac{\pi e^2\hbar}{2\epsilon_0 m_ec}\frac{{\rm d}f}{{\rm d}E}
\label{eqn-crosssection}
\end{equation}
with the electron charge $e$ and mass $m_e$, the reduced Planck constant $\hbar$, the speed of light $c$ and the electric constant $\epsilon_0$. For an EUV wavelength of 17.7 nm (photon energy 70 eV), the photoionization cross section of helium is $\sim$1 Mb ($10^{-18}~\rm cm^2$) \cite{samson_precision_2002}.

\begin{figure}[h]%
\centering
\includegraphics[width=0.65\textwidth]{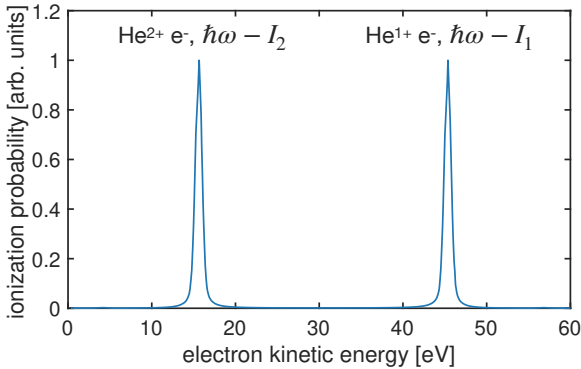}
\caption[Electron kinetic energy spectrum of two-photon double ionization (TPDI) of helium]{Electron kinetic energy spectrum of two-photon double ionization (TPDI) of helium by an EUV pulse at $\hbar\omega=70$ eV and 4.5 fs pulse duration. The photoelectron energy for $\rm He^{1+}$ electron is $\hbar\omega-I_1=45.4$ eV, higher than $\hbar\omega-I_2=15.6$ eV for $\rm He^{2+}$ electrons. The plot is made using the TDSE simulation reported in \cite{feist_electron_2009}.}
\label{fig-TPDI}
\end{figure}

After being ionized, the electrons are ejected along the electric field vector of the photon beam with an energy equal to the difference in photon energy and the ionization energy (excess energy). The free electron distribution functions can thus be like the line shape function of the photon pulse. For the case of He, when a single frequency photon beam is used, the distribution can be approximated as two delta functions at the excess energy \cite{feist_electron_2009,feist_probing_2009} as shown in Fig. \ref{fig-TPDI}.

As we will see next, contrary to the high field tunnel ionization regime, single photon ionization will yield $\rm He^{1+}$ electrons with a higher energy than $\rm He^{2+}$ electrons (see the graphic in Fig. \ref{fig-TPDI}). Nevertheless, the relative streaming between these two electron groups can lead to TSI and CFI \cite{bychenkov_distinctive_2006}.

\subsection{High-field photoionization in the tunneling regime}\label{sec-tunneling}
\begin{figure}[h]%
\centering
\includegraphics[width=0.95\textwidth]{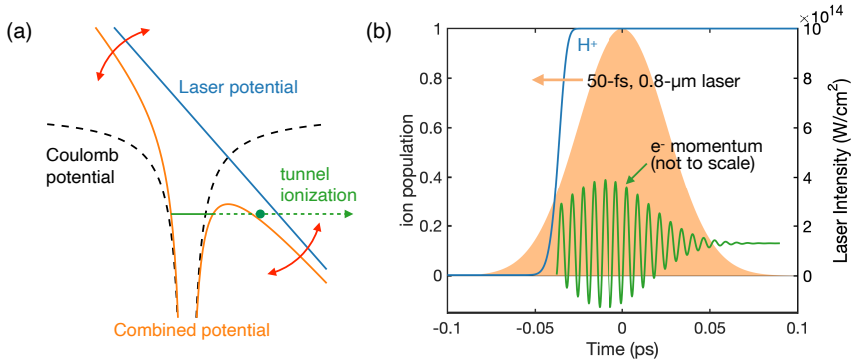}
\caption[Tunnel ionization and residual energy of optical-field ionized electrons.]{Tunnel ionization and residual energy of photoionized electrons. (a) Illustration of tunnel ionization process. The Coulomb potential of hydrogen atom (dashed black line) is modified by the presence of an intense laser field (blue line) which leads to the formation of a combined potential (orange line) that allows the electron (represented by the green dot) to tunnel through the finite barrier. Ionization calculation of hydrogen gas by an ultrashort, circularly polarized laser pulse with peak intensity of $10^{15}~\rm W/cm^2$ and pulse duration of 50 fs (FWHM). The intensity profile of the laser is shown in orange shaded area. The ion population ($\rm H^+$) calculated using the ADK is shown by the blue line. The momentum of a representative electron is sketched in green curve, showing a nonzero residual momentum after the laser has left.}
\label{fig-ionization}
\end{figure}

In this section, we present a second method for initializing the EVDs using an ultrashort but intense laser pulse to ionize underdense neutral gas in the high field, long wavelength limit- often called the tunnel ionization regime. This regime is defined by the Keldysh parameter \cite{keldysh_ionization_1965}, $\gamma\equiv\omega\sqrt{2I_p}/E_0\ll1$ where $I_p$ is the ionization potential of the atom, $\omega$ the frequency and $E_0$ is the electric field of the laser pulse. This regime is easily reached by tightly focusing femtosecond laser pulses with a few tens of mJ energy. The tunnel ionization of a hydrogen atom is illustrated schematically in Fig. \ref{fig-ionization} (a) at one instance in time. In the presence of an intense laser field, the Coulomb potential (black dashed line) of a hydrogen atom is significantly asymmetrically altered by the laser field (blue line), resulting in a combined potential (orange curve) with a suppressed barrier on one side from which the electron (green dot) can tunnel through before the field changes sign and become a free electron \cite{augst_tunneling_1989}. After that, the electron is accelerated (perpendicular to wave vector $\bf k$ of the laser and displaced in the direction of $\bf k$) because of the $\bf v\times B$ force exerted by the magnetic field of the laser field and released with energy equal to the vector potential of the laser at the instant of ionization. The electron velocity is in the direction of the laser electric field when the laser is gone.

Even though tunneling is a quantum mechanical concept, a semi-classical model has been successful at predicting many  high field phenomena such as high harmonic generation \cite{corkum_plasma_1993}, attosecond train generation \cite{krausz_attosecond_2009}, angular distribution of photoelectrons \cite{reid_photoelectron_2003}, and the energy spectrum of ionized electrons \cite{lafon_electron_2001,rae_possible_1992}.

Figure \ref{fig-ionization}(b) depicts an example calculation of the ionization of hydrogen gas by a 0.8 $\rm \mu m$, circularly polarized laser with a peak intensity of $10^{15}~\rm W/cm^2$ and a duration of 50 fs laser using the Ammosov-Delone-Krainov (ADK) model \cite{ammosov_tunnel_1987}. As the laser intensity increases above the ionization threshold ($\sim10^{14}\rm W/cm^2$), the ion population grows rapidly, as shown by the blue curve. Full ionization occurs in about three laser cycles ($<$10 fs). The green line sketches the momentum evolution of a representative electron, which oscillates when the laser is present and ultimately retains energy corresponding to the finite residual vector potential at the instant when it was ionized only after the laser has passed.

To calculate the transverse momentum of the electron in the general case of arbitrary $a_0\equiv eA/m_ec^2$, we apply the conservation of the canonical momentum, yielding the following expression,
\begin{equation}
p_\perp(t)-\frac{e}{c}A_\perp(t)=\rm Const
\label{eqn-canonical}
\end{equation}
where $p_\perp$ is the transverse kinetic momentum of the electron, $A_\perp$ represents the transverse vector potential of the laser (i.e., perpendicular to the laser's propagation direction), and $\rm Const$ is the constant of motion. Assuming the electron is at rest at $t_0$, the residual transverse momentum after the laser is gone is given by $p_\perp=-e/cA_\perp(t_0)$ \cite{huang_transient_nodate}.

\begin{figure}[h]%
\centering
\includegraphics[width=0.95\textwidth]{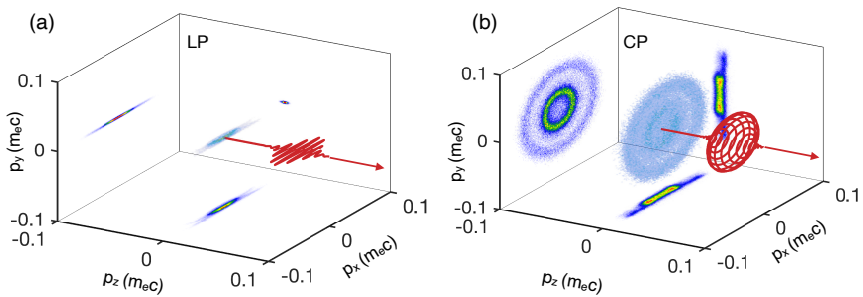}
\caption[Highly anisotropic electron velocity distributions of optical-field ionized helium plasmas.]{Highly anisotropic electron velocity distributions of optical-field ionized helium plasmas. (a) Momentum space of a helium plasma ionized by a linearly polarized laser. The 2D projections of the electron distribution are shown, with the red oscillating line representing the electric field of the laser propagating in the direction indicated by the arrow. (b) Similar plot for the case of circular polarization. The EVDs shown in both (a) and (b) are $\sim$130 fs after the onset of the ionization of the first helium electron ($\sim$87 fs after the peak of the laser pulse has passed).}
\label{fig-an-anisotropicEVD}
\end{figure}

This analysis indicates that in optical-field ionization, the electron velocity distribution is determined by the instantaneous vector potential of the laser when ionization happens. Subsequently, the initial EVDs of the plasma electrons can be altered simply by changing the laser parameters, such as polarization \cite{huang_initializing_2020}. Figure \ref{fig-an-anisotropicEVD} displays two examples of electron velocity distributions obtained from self-consistent PIC simulations \cite{zhang_probing_2020}. In these simulations, we used laser pulses with a duration of 60 fs (FWHM), a spot size of 8 $\rm\mu m$, and a peak normalized vector potential of 0.2 to ionize helium gas with an initial density of $5\times10^{18}~\rm cm^{-3}$. The simulations modeled the ionization process using the ADK model and followed the subsequent motion of the electrons self-consistently.

In Fig. \ref{fig-an-anisotropicEVD}(a), the plasma ionized by a linearly polarized laser is hottest along the laser polarization direction and colder in the other two orthogonal directions as expected, with subtle fork structures arising from the different ionization potentials of the two helium electrons. Also note that although ionization of the first and the second He electron occur within 2-3 laser cycles there is a temporal difference of additional 20 fs between the peaks of two ionization rates for the intensity risetime of our pulse. The helium atoms are fully ionized before the laser reaches peak intensity. Thus, time $t=0$ at which the overall EVD is initialized is delayed from the onset of the ionization of the first He electron by $\sim$130 fs during which electrons can diffuse towards the axis of the laser beam. This explains why electrons have partially filled the region enclosed by the inner ring in the $p_x$ vs $p_y$ projection of Fig \ref{fig-an-anisotropicEVD}(b).

Along the laser polarization direction, the distribution can be approximated by the sum of two Maxwellian distributions with different temperatures. In contrast, Fig. \ref{fig-an-anisotropicEVD}(b) shows a significantly different velocity distribution for the circular polarization case. Instead of peaking at zero, the distribution exhibits two distinct rings corresponding to the two groups of $\rm He^{1+}$ and $\rm He^{2+}$ electrons. This is because the electric field of the circularly polarized laser only rotates in the transverse plane but does not vanish, leading to all electrons acquiring nonzero momentum in the radial direction. The transverse distribution is extremely non-thermal. In both cases, the distributions are highly anisotropic, with the temperature along the laser electric field direction much higher than in the other directions, making them suitable for triggering kinetic instabilities such as streaming, current filamentation, and Weibel instability.

The momentum distribution of photoelectrons (or ions) produced by strong field ionization has been studied extensively in atomic physics \cite{okunishi_experimental_2008,yuan_photoelectron_2016,maurer_molecular-frame_2012}. For example, velocity map imaging has been used to accurately measure the momentum distribution of photoemitted electrons or ions \cite{eppink_velocity_1997,dorner_cold_2000,liu_strong-field_2008}. However, the application of this technique is limited to extremely low-density gases where interactions between electrons are negligible (i.e., the single-atom regime). Therefore, this technique is not applicable for the relatively dense plasmas discussed here. Consequently, a different approach is needed to measure/infer the electron velocity distribution in the bulk of the plasma.

\subsection{Thomson scattering measurement of electron velocity distributions}
Thomson scattering is a well-established method for diagnosing plasma parameters and has been widely used to measure the density and temperature of laboratory plasmas \cite{sheffield_plasma_2010}. The Thomson scattering parameter, $\alpha\equiv(k\lambda_{De})^{-1}$, separates the collective and thermal scattering regimes. Here, $k=2\pi/\lambda$ is the scattering wavenumber, which is determined by the incident and scattered light wavenumbers and the scattering geometry, and $\lambda_{De}$ is the Debye length of the plasma. In the collective regime, Landau damping is negligible ($\alpha>1$, or equivalently $k\lambda_{De}<1$), most electrons within a Debye sphere radiate in phase, leading to strong correlation between (Thomson) scattered photon spectrum via the electron susceptibility. On the other hand, in the thermal scattering regime, Landau damping is very large ($\alpha\ll1$), electrons within a Debye sphere radiate randomly in phase and do not correlate with each other. As a result, the spectral density function reduces to $S(k,\omega)\simeq2\pi/k f_e(\omega/k)$, which implies that the scattered light spectrum maps the electron velocity distribution. This property is the basis of using thermal Thomson scattering to infer the electron velocity distribution.

\begin{figure}[h]%
\centering
\includegraphics[width=0.95\textwidth]{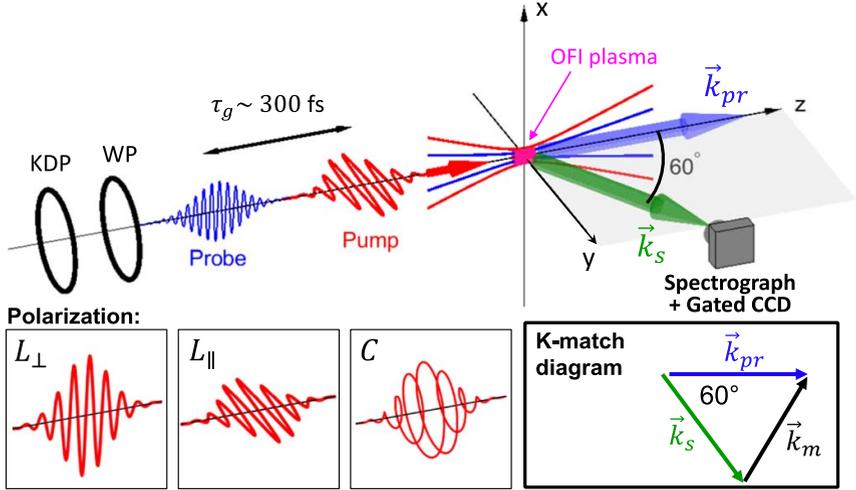}
\caption[Schematic of the collinear Thomson scattering (TS) experiment.]{Schematic of the collinear Thomson scattering (TS) experiment. The 800 nm pump beam generates photoionized plasmas that are probed by a collinear 400 nm TS beam at a fixed delay. The linear polarization of the probe can be adjusted to be either perpendicular to the scattering plane ($L_\perp$), parallel to the scattering plane ($L_\parallel$), or circularly polarized (CP). The k-matching diagram shows the vector $k_m$ probed in Thomson scattering. KDP refers to the KDP crystal, and WP refers to the half-wave plate for linear polarization or quarter-wave plate for CP.}
\label{fig-TS-EVD-setup}
\end{figure}

Experiments that measure the initial electron velocity distribution of tunnel ionized plasmas have been done at UCLA \cite{huang_initializing_2020}, with the experimental setup illustrated in Fig. \ref{fig-TS-EVD-setup}. The laser pulse delivered by a Ti:Sapphire laser was split into two pulses. The 800-nm, 50-fs, 10-mJ pump pulse with controllable polarization was focused to a spot size of $w_0 = 8~\rm \mu m$, giving a peak intensity of approximately $10^{17}~\rm W/cm^2$. The less intense probe pulse (1 mJ, 90 fs) was generated by frequency doubling the original pulse using a KDP crystal and then recombined with the pump pulse collinearly, with a fixed delay of about 300 fs. The scattered light was collected at a $60^\circ$ angle with respect to the propagation direction of the incident pulses by an imaging spectrometer defining a scattering volume of $20 \times 20 \times 20~\rm\mu m^3$.

\begin{figure}[h]%
\centering
\includegraphics[width=0.95\textwidth]{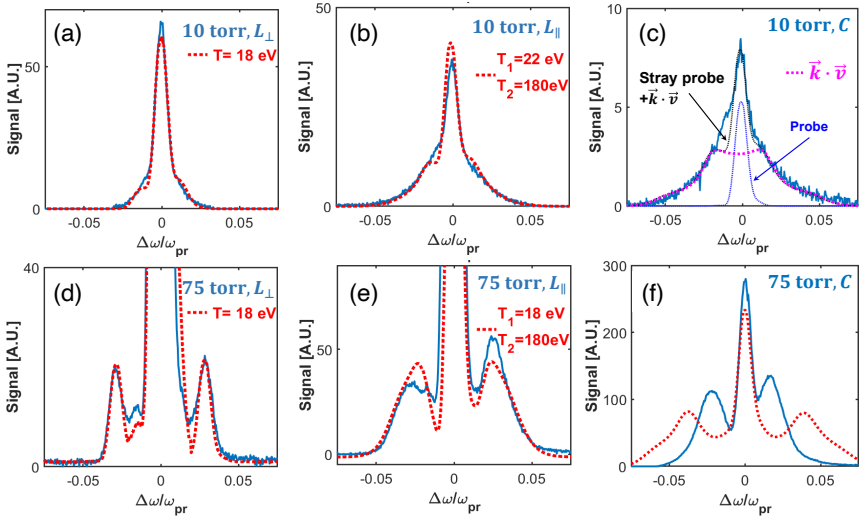}
\caption[Thomson scattering measurement results.]{Thomson scattering measurement results. The blue solid lines represent the measured Thomson scattering spectra obtained by averaging over 200 consecutive shots to improve the signal-to-noise ratio. Top row (a)-(c) show results for low density ($n_e\approx7\times10^{17}~\rm cm^{-3}$) while the bottom row (d)-(f) show results for high density ($n_e\approx5\times10^{18}~\rm cm^{-3}$). The red dashed lines indicate the best fit using Maxwellian velocity distributions. Panels (a) and (d) show the results for linear polarization perpendicular to the scattering plane ($L_\perp$) using a single-temperature ($T_e=18$ eV) Maxwellian distribution. Panels (b) and (e) present the results for linear polarization parallel to the scattering plane ($L_\parallel$) showing the best fit using a two-temperature Maxwellian distribution. Panels (c) and (f) show the results for circular polarization, where the magenta dashed line in (c) represents the Doppler shift spectrum, calculated using experimental $\bf k$ and velocity distribution $f({\bf v})$ from 3D PIC simulation. The blue dotted line indicates the spectrum of the probe, and the black dotted line shows the synthetic spectrum that can be compared with the measurement. The red dashed line in (f) represents the calculated spectrum using a distribution with two pairs of counter-drifting Maxwellian streams, deduced from the electron velocity distribution obtained from the EVD in PIC simulation.}
\label{fig-TS-EVD-results}
\end{figure}

The blue solid lines in Fig. \ref{fig-TS-EVD-results} show the measured Thomson scattering spectra for three different polarization configurations and two different static fill pressures. The top row shows the results for a relatively low density of $n_e\approx7\times10^{17}~\rm cm^{-3}$, corresponding to a characteristic time scale of $2\pi/\omega_p^{-1}\approx130$ fs. The red dashed lines in Fig. \ref{fig-TS-EVD-results}(a) and (b) represent the best fits to the experimental data. For the $L_\perp$ case shown in Fig. \ref{fig-TS-EVD-results}(a), the data can be well fitted using a single-temperature Maxwellian distribution with $T_e=18\pm2$ eV. In the $L_\parallel$ case in Fig. \ref{fig-TS-EVD-results}(b), however, a two-temperature distribution with $T_{e1}=22$ eV and $T_{e2}=180$ eV is needed to best fit the data. This is because the two groups of electrons have different residual momenta due to their different ionization potentials. The difference in the velocity distributions of these two groups of electrons (i.e., the temperatures) is predominant in the laser polarization direction. In contrast, the temperature difference in the orthogonal direction is too small, such that a single-temperature Maxwellian distribution fits the data. Self-consistent 3D PIC simulation shows that the plasma temperature in the direction orthogonal to both the propagation and polarization direction of the pump is 12 eV, which agrees reasonably well with the measured (best fit) $18\pm2$ eV for the $L_\perp$ case. The measured $L_\parallel$ temperatures, $T_{e1}=20\pm2$ eV and $T_{e2}=180\pm20$ eV also reasonably agree with those extracted from simulation that corresponds to $\rm He^{1+}$ and $\rm He^{2+}$ electrons projected to the scattered $\bf k$ direction, 45 eV and 160 eV.

Circularly polarized pump is expected to generate a much hotter and nonthermal electron velocity distribution, as shown in Fig. \ref{fig-TS-EVD-results}(c). This spectrum cannot be fitted using Maxwellian distributions but can be well fitted using the Doppler frequency shift $\Delta\omega={\bf k\cdot v}$, where $\bf k$ is the probed wavevector and $f({\bf v})$ is the electron velocity distribution obtained from self-consistent 3D PIC simulation [see Fig. \ref{fig-an-anisotropicEVD}(b)]. The electron velocity distribution contains two distinct ring structures in the velocity space, and the projection of these structures along the $\bf k$ direction exhibits a small dip at $\Delta\omega=0$ [see the magenta line in Fig. \ref{fig-TS-EVD-results}(c)].

The results for a higher density, $n_e\approx5\times10^{18}~\rm cm^{-3}$, are presented in the bottom row of Fig. \ref{fig-TS-EVD-results} for comparison. In the linear polarization cases, the measured spectra can still be fitted with one- ($L_\perp$) or two-Maxwellian distributions ($L_\parallel$), indicating that collective effects such as kinetic instabilities have not significantly altered the distribution. In contrast, the measured spectrum in the circular polarization case shown in Fig. \ref{fig-TS-EVD-results}(f) exhibits two side peaks, which correspond to collective features but cannot be fitted using the Doppler shift term nor a two-temperature Maxwellian distribution. As an illustration, the calculated spectrum using a distribution with two pairs of counter-drifting Maxwellian streams (drift velocities of $\pm0.015c$ and $\pm0.046c$, with widths of 87 and 79 eV and density ratio of 4:1) deduced from the EVD in PIC simulations is shown by the red dashed line in Fig. \ref{fig-TS-EVD-results}(f). However, the peak locations in the calculated spectrum do not match the measurements. Measurements taken under different densities also reveal that the peak locations (frequency shift) remain unchanged, indicating that these peaks do not correspond to the normal electron features in collective Thomson scattering that should follow the Bohm-Gross frequency shift $\omega_{BG}\simeq(\omega_{pe}^2+3k_BT_e k^2/m_e)^{1/2}$. This failure to predict even the peak of the Thomson scattered photon correctly- when CP pump beam produces a high-density plasma- suggests that some other mechanism(s) are at work that dominate the scattered photon spectral distribution. In Sec. \ref{sec3}, we will discuss different kinetic instabilities that can grow in photoionized plasmas and their role in thermalizing and/or isotropizing the plasma and in Sec. \ref{sec4-hierarchy} we show how the Thomson scattered light spectra can enable us to infer the real (frequency) and imaginary (linear growth rate) frequency parts of these instabilities.

\subsection{EVD by design}
\begin{figure}[h]%
\centering
\includegraphics[width=0.95\textwidth]{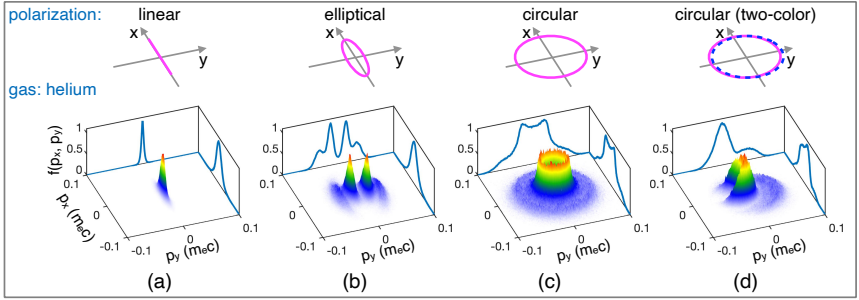}
\caption[Examples of controllable initialization of EVDs.]{Examples of controllable initialization of EVDs of helium plasma using lasers with various polarizations and/or wavelengths. Panels (a)-(c) display the distribution in the transverse plane, $f(p_x, p_y)$, for three different polarizations, i.e., linear, elliptical, and circular, with the laser wavelength fixed at 800 nm. Panel (d) presents the EVD for a two-color ionization case, where two overlapping lasers with 800 and 400 nm wavelength are used to ionize the helium gas.}
\label{fig-EVD-by-design}
\end{figure}

We have demonstrated the effective initialization of nonthermal and highly anisotropic electron velocity distributions using photoionization with linear or circular polarizations. As indicated by Eqn. \ref{eqn-canonical}, it is possible to initialize plasmas with more sophisticated electron velocity distributions in a controllable manner by engineering the laser fields. Some examples are shown in Fig. \ref{fig-EVD-by-design}, with panel (a) and (c) show the transverse momentum distribution $f(p_x, p_y)$ of helium plasma ionized by a linearly or circularly polarized laser propagating along the $z$ direction, which have been verified in experiments using Thomson scattering.

Upon changing to elliptical polarization, as shown in Fig. \ref{fig-EVD-by-design}(b), the distributions of $\rm He^{1+}$ and $\rm He^{2+}$ electrons separate along the $y$ direction. The two higher peaks with smaller $p_y$ correspond to $\rm He^{1+}$ electrons, whereas the outer two lower peaks are due to $\rm He^{2+}$ electrons. Each individual peak is still hotter in the $x$ direction, along which the laser field is stronger. Another example is displayed in panel (d), where helium gas is ionized by a circularly polarized two-color pulse (superposition of a 0.8 $\rm\mu m$ laser pulse and its second harmonic). In this case, the symmetry of the momentum distribution is broken due to the asymmetric waveform of the ionizing laser. The velocity distribution in the $y$ direction clearly shows a bump-on-tail shape, which can trigger the growth of TSI. These examples demonstrate the feasibility of controlling the plasma EVD by engineering the laser waveform, which can be done by changing the laser polarization or by combining different wavelengths, as shown here and in the literature \cite{zhang_laser-sub-cycle_2014}, or by using more sophisticated laser pulses such as those having temporal chirp, orbital angular momentum or even structured light \cite{pierce_arbitrarily_2023,rubinsztein-dunlop_roadmap_2016}, as shown in the single-atom regime \cite{smaxwell_manipulating_2021,fang_probing_2022}.

\section{Kinetic instabilities in photoionized plasmas}\label{sec3}
In this section, we discuss how the nonthermal and highly anisotropic electron velocity distribution described earlier can lead to the growth of various kinetic instabilities. The TSI and CFI are triggered by interpenetrating electron streams. As the TSI grows, it converts the excess kinetic energy into electric energy by exciting electrostatic modes in the plasma. In contrast, the CFI and WI convert kinetic energy into magnetic energy by amplifying magnetic fields from noise. The growth of these instabilities can be analyzed by solving the dispersion relation. A detailed analysis of a beam-plasma system is given by Bret \cite{bret_multidimensional_2010}.

\subsection{Streaming instability}\label{sec3-streaming}
The dispersion relation for longitudinal plasma waves in the one-dimension case is derived by combining the linearized Vlasov equation and Maxwell’s equations as
\begin{equation}
1- \frac{\omega_p^2}{n_0k^2}\int\frac{f_0(v)}{(v-\omega/k)^2}{\rm d}v=0.
\label{eqn-disper-streaming1}
\end{equation}
Assuming a Maxwellian distribution, the dispersion relation for non-relativistic two-stream (or multi-stream) instability in 1D can be derived as
\begin{equation}
1- \sum_j \frac{\omega_{pj}^2}{2k^2 v_{th, j}^2} Z^\prime\left( \frac{\omega-kv_{j}}{\sqrt{2}k^2v_{th,j}^2} \right)=0
\label{eqn-disper-streaming}
\end{equation}
where $Z^\prime(\xi)=-2[1+\xi Z(\xi)]$ and $Z(\xi)=\pi^{-1}\int_{-\infty}^{\infty} e^{-t^2}/(t-\xi)dt$ is the plasma dispersion function. The parameters $v_{0j}$, $v_{th,j}$, and $\omega_{pj}$ represent the drift and thermal velocity and the plasma frequency for the $j^{th}$ stream, respectively. Numerical solutions for Eqn. \ref{eqn-disper-streaming} can be obtained using a dispersion relation solver such as \cite{xie_bo_2019} to get $\omega=\omega_r+i\omega_i$, where $\omega_r$ is the real oscillating frequency of the charge perturbation and $\omega_i$ denotes the growth rate of the instability.
\begin{figure}[h]%
\centering
\includegraphics[width=0.95\textwidth]{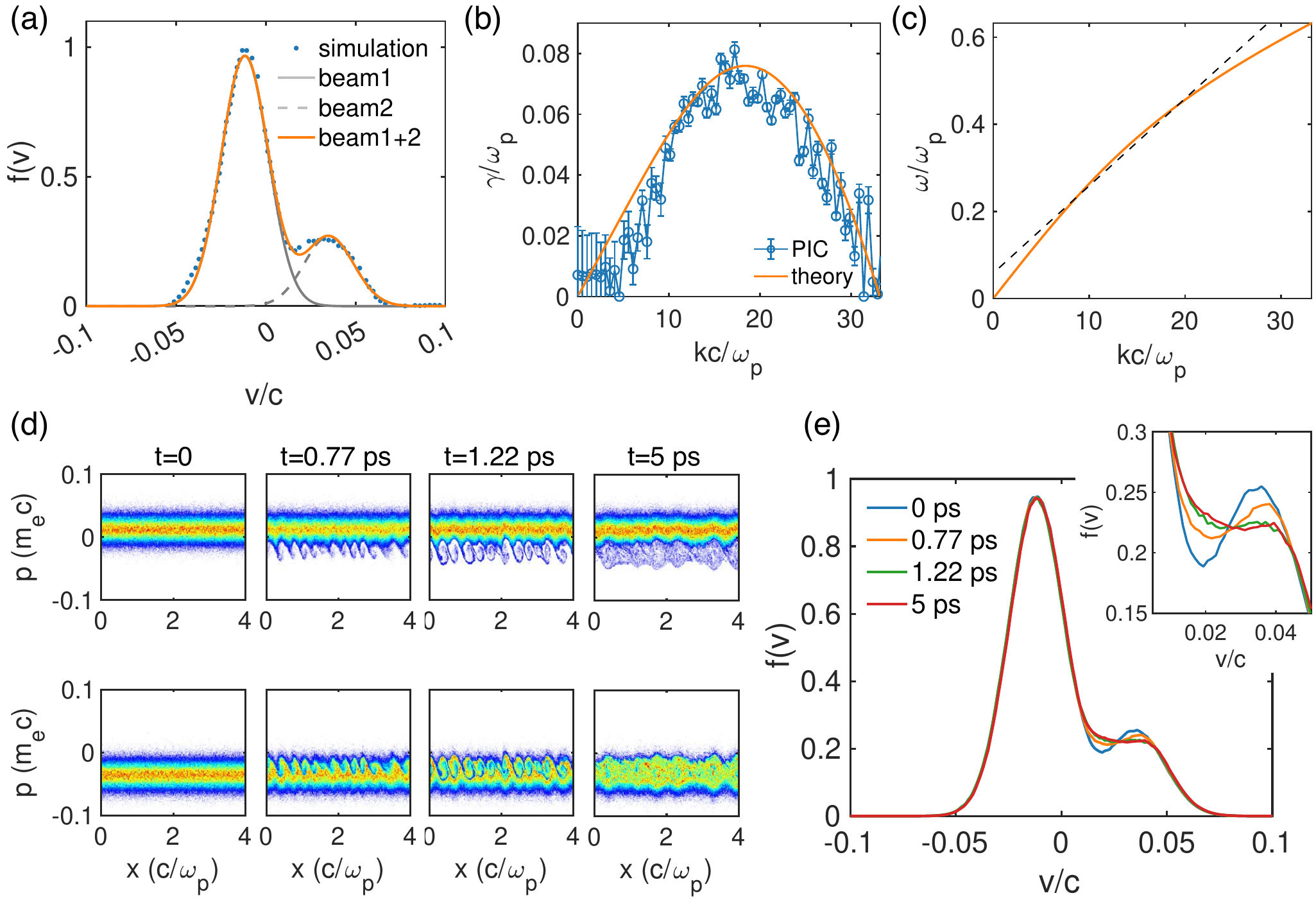}
\caption[Simulation results of two-stream instability driven by a bump-on-tail distribution.]{Simulation results of two-stream instability driven by a bump-on-tail distribution. (a) The electron velocity distribution (EVD) deduced from particle-in-cell (PIC) simulation [Fig. \ref{fig-EVD-by-design}(d)] and the best fit using Maxwellian distributions. (b) Growth rate of the two-stream instability predicted by kinetic theory (orange line) and obtained from PIC simulation (blue line). (c) Real (oscillating) frequency of the instability calculated using kinetic theory. (d) Phase spaces of the two streams as a function of simulation time. (e) Evolution of the EVD. The inset highlights the flattening of the bump in the original distribution due to the growth of the instability.}
\label{fig-two-stream}
\end{figure}

The electrostatic oscillation is subject to Landau damping/growth, which leads to energy exchange between the two streams and flattening of the velocity distribution. Figure \ref{fig-two-stream} summarizes the results of a 1D simulation that exemplifies the characteristics of the TSI. The electron velocity distribution is inferred from a photoionized plasma example that uses a two-color laser pulse, as shown in Fig. \ref{fig-EVD-by-design}(d). This distribution can be approximated by two beams with different densities, drifting and thermal velocities [beam 1 and 2 in Fig. \ref{fig-two-stream}(a)]. The growth rate and real frequency are determined by solving the dispersion relation (Eqn. \ref{eqn-disper-streaming}) and plotted in Fig. \ref{fig-two-stream}(b) and (c) as orange lines. The calculated growth rate exhibits a broad unstable spectrum that covers $0<k<33~\omega_p/c$ and peaks at $18~\omega_p/c$. A 1D particle-in-cell (PIC) simulation is carried out using the code Osiris\cite{goos_osiris_2002} and the beams in Fig. \ref{fig-two-stream}(a) with densities $n_1$ and $n_2$ so that $n_1+n_2=n_p$ and a fixed ion background. The total density of the beams is $n_p=10^{18}~\rm cm^{-3}$ with $n_1=0.7731n_p$, $v_{1y}=-0.01174c$, $v_{1,th,y}=0.01313c$, $v_{1,th,x,z}=0.002c$ and $n_2=0.2269n_p$, $v_{2y}=0.03509c$, $v_{2,th,y}=0.01379c$, $v_{2,th,x,z}=0.002c$. The simulation is spatially resolved in the $x$ direction with cell size of $1/64 c/\omega_p$ and 4096 particles per cell.

From the simulation, the growth rate of the electric field $E_x$ is obtained and plotted as the blue line, which agrees well with the kinetic theory calculation within the uncertainty range of the PIC simulation due to finite resolution in both space and time. The phase velocity of the excited wave is extracted from Fig. \ref{fig-two-stream}(c) as $v_\phi\approx 0.02c$ [see the dotted line in Fig \ref{fig-two-stream} (c)], which resonates with the velocity of electrons that contribute to the dip in the distribution function, thereby leading to efficient damping of the wave and flattening of the distribution function. The top 4 frames of Fig. \ref{fig-two-stream} (d) show the phase space of $\rm He^{2+}$ electrons whereas the bottom 4 frames show the phase space of $\rm He^{2+}$ electrons. At 0.77 ps the streaming between these two distributions has produced electrostatic wave that has the highest growth rate (b). This wave traps electrons belonging to both species decelerating the $\rm He^{2+}$ and accelerating $\rm He^{1+}$ electrons and flattening out the bump on tail distribution shown in Fig. \ref{fig-two-stream} (e).

\subsection{Current filamentation instability}\label{sec3-CFI}
The same two-stream configuration is also susceptible to the CFI. The dispersion relation governing the 1D CFI for beams with Maxwellian distributions is given by
\begin{equation}
1- \frac{\omega_p^2+c^2k_z^2}{\omega^2} - \sum_j \frac{\omega_{pj}^2}{\omega^2}\frac{v_{jy}^2+v_{th,jy}^2}{2v_{th, jz}^2} Z^\prime\left( \frac{\omega}{\sqrt{2}k_z v_{th,jz}} \right)=0
\label{eqn-disper-filamentation}
\end{equation}
where $\omega_{pj}$, $v_{jy}$, $v_{th, jy}$, and $v_{th, jz}$ represent the plasma frequency, drift velocity, and thermal velocities in the $y$ and $z$ directions for the $j^{th}$ beam, respectively. The derivation neglects the ion response by assuming a fixed ion background. Both TSI and CFI have a range of $k$s that are unstable, with TSI being a longitudinal mode with electron density perturbation whereas CFI being a primarily transverse mode with electron current density perturbation. For transversely cold beams with $v_{th, jz}=0$, the dispersion relation simplifies to
\begin{equation}
1- \frac{\omega_p^2+c^2k_z^2}{\omega^2} - \sum_j \frac{\omega_{pj}^2}{\omega^2} (v_{jy}^2+v_{th,jy}^2)=0
\label{eqn-disper-filamentation-cold}
\end{equation}

\begin{figure}[h]%
\centering
\includegraphics[width=0.95\textwidth]{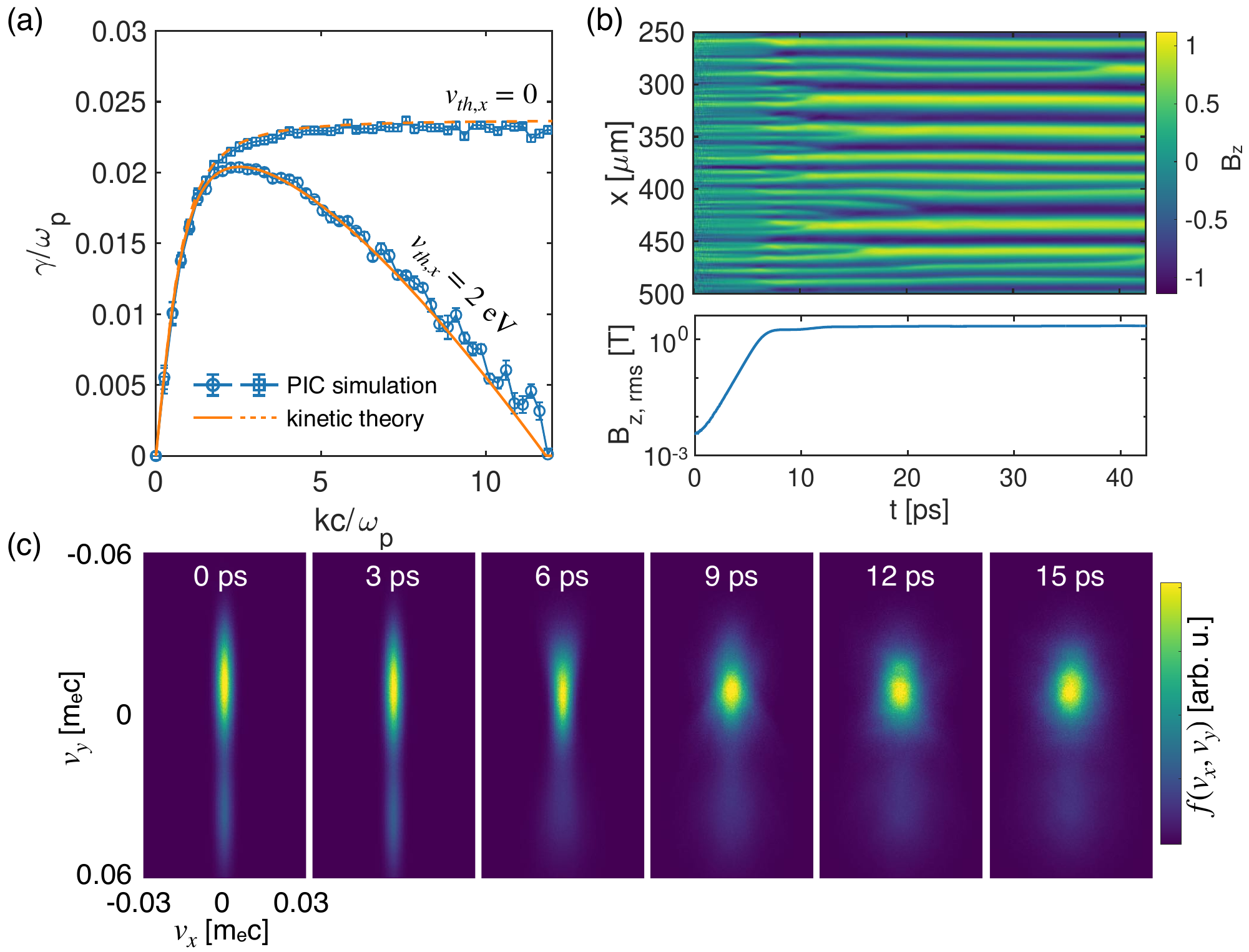}
\caption[Simulation of current filamentation instability.]{Simulation of current filamentation instability. (a) The $k$-resolved growth rates of the current filamentation instability for transversely cold ($v_{th,x}=0$) and thermal ($v_{th,x}=2$ eV) beams obtained from 1D PIC simulations. The simulation results are in excellent agreement with the predictions of kinetic theory (orange lines). (b) Amplification of the magnetic field $B_z$ due to the continuous merging of plasma currents. Each slice of $B_z$ field is normalized to the instant peak. The bottom panel shows the exponential growth of the $B_z$ field and the subsequent saturation. The plasma period 
$\omega_p^{-1}$ is 0.018 ps. (c) The transverse velocity phase spaces at representative times, revealing the isotropization of the plasma during the evolution of the current filamentation instability.}.
\label{fig-CFI}
\end{figure}

We performed 1D simulations to illustrate the growth of CFI using the bump-on-tail distribution shown in Fig. \ref{fig-two-stream}(a). The two streams drift along the $y$ direction, and the magnetic fields along $z$ are resolved spatially in the $x$ direction. Two simulations were conducted: one with streams cold in the $x$ direction ($v_{th,x}=0$) and the other with a small transverse temperature ($v_{th,x}=2$ eV) such that the largest $k$ that can grow is limited due to the balancing between the transverse velocity spread and the bunching of microscopic plasma currents with short wavelengths. The $k$-resolved growth rates of $B_z$ were deduced from these PIC simulations and are plotted in Fig. \ref{fig-CFI}(a). In both cases, the simulated growth rates are in excellent agreement with the prediction of the kinetic theory (solutions of Eqn. \ref{eqn-disper-filamentation} and \ref{eqn-disper-filamentation-cold}). The growth rate for the cold-beam case ($v_{th,x}=0$) extends to large $k$ with a constant growth rate. In contrast, the thermal-beam case has a largest unstable $k=\sqrt{A}\approx11.8~\omega_p/c$ for the given beam parameters, which is also validated in the PIC simulation.

The bunching process of the plasma currents is illustrated in Fig. \ref{fig-CFI}(b) by showing the spatiotemporal evolution of the $B_z$ field, which is $\pi$ out of phase with the current $J_y$. The continuous bunching of plasma currents occurs throughout the linear growth period for $t<8$ ps, which amplifies the magnetic fields by approximately three orders of magnitude. Once the instability reaches saturation ($\sim$10 ps), the magnetic fields become quasi-static and evolve much slower on the timescale of tens of picoseconds. Nevertheless, merging of current filaments still occurs, albeit at a lower rate.

These self-generated magnetic fields, begin to bend the trajectory of electrons, leading to the isotropization of the plasma. This phenomenon is depicted in Fig. \ref{fig-CFI}(c), where velocity space snapshots at representative times are presented. Initially, the transverse spread of the beams ($v_x$) in the first two snapshots is relatively small and remains unchanged due to the weak magnetic fields. Between 6 and 12 ps, the transverse spread of the beam rapidly increases as more electrons are deflected by the fast-growing magnetic fields. Thereafter, the phase space of the plasma evolves at a slower rate.

\subsection{Thermal Weibel instability}\label{sec3-weibel}
The counter-propagating beams in the example shown in Fig. \ref{fig-CFI} can drive the CFI. However, the mechanism of microscopic plasma current bunching does not require the presence of beams and can also happen in a stationary plasma with temperature anisotropy due to the thermal Weibel instability \cite{weibel_spontaneously_1959}. WI is equivalent to the limiting case of CFI with $v_{jy}=0$ in Eqn. \ref{eqn-disper-filamentation}. The dispersion relation for the thermal Weibel instability driven by bi-Maxwellian EVDs is given by
\begin{equation}
1- \frac{c^2k_z^2}{\omega^2} + \frac{\omega_p^2}{\omega^2}\left[A+(A+1)\frac{\omega}{\sqrt{2}k_z v_{th,z}} Z\left(\frac{\omega}{\sqrt{2}k_z v_{th,z}} \right) \right]=0
\label{eqn-disper-weibel}
\end{equation}
where $A\equiv T_{hot}/T_{cold}-1=v_{th,y}^2/v_{th,z}^2-1$ is the temperature anisotropy and $v_{th,y}$ and $v_{th,z}$ are the thermal velocities in the perpendicular and parallel directions to the background magnetic field, respectively. In an infinite plasma, the instability grows from current fluctuations inherent in the temperature anisotropy with a broad $k$-spectrum, $0<k<\sqrt{A}~\omega_p/c$, indicating the excitation of many modes simultaneously, each with an effective growth rate. As the temperature anisotropy decreases, the $k$-spectrum narrows to a peak due to the coalescence of plasma currents and the amplification of magnetic fields. Like the CFI, once the quasi-single mode is formed, the magnetic field maintains its topological structure for many plasma periods, $\gg\omega_p^{-1}$, as shown in Fig. \ref{fig-CFI}(b). Indeed the final topology of the magnetic field structure is similar in both cases but the orientation, periodicity and the magnitude of the magnetic field can be very different depending on the orientation and the magnitude of the temperature anisotropy and in the case of CFI the direction of the drifting streams.

\begin{figure}[h]%
\centering
\includegraphics[width=0.95\textwidth]{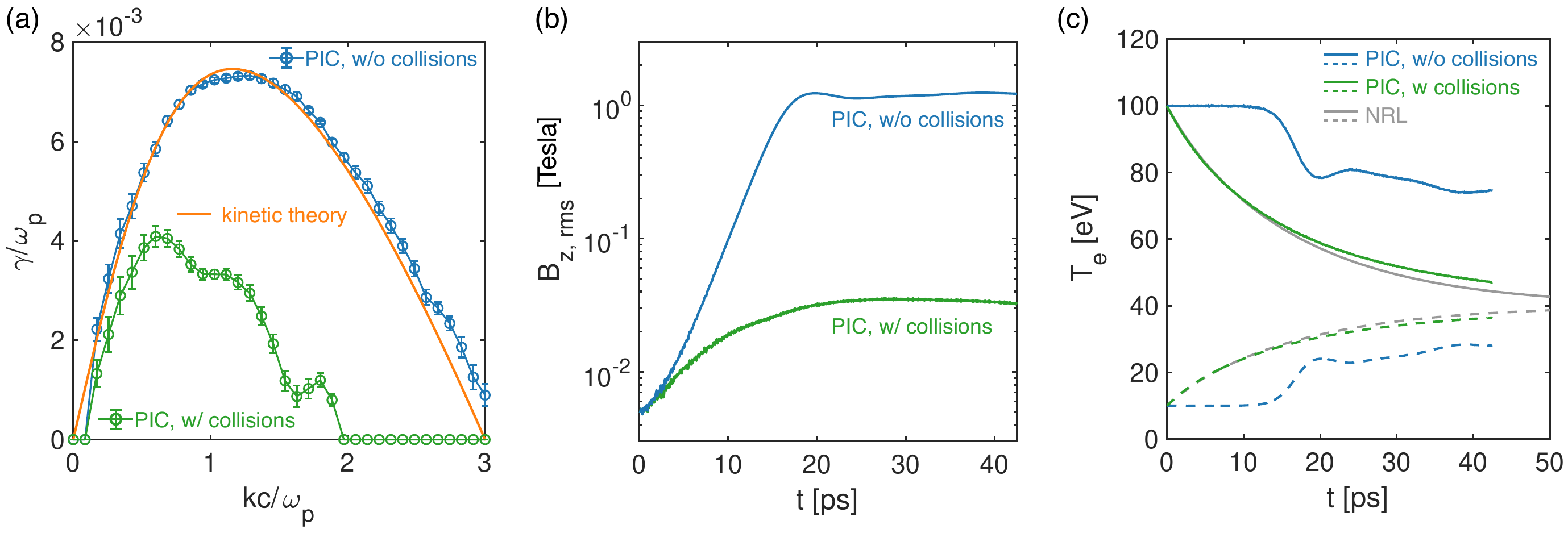}
\caption[Simulation of thermal Weibel instability and effects of collisions.]{Simulation of thermal Weibel instability and effects of collisions. (a) The k-resolved growth rate of the Weibel instability obtained from 1D PIC simulations with (green) and without (blue) binary collisions. The collisionless simulation results agree with the theoretical prediction from the dispersion relation (Eqn. \ref{eqn-disper-weibel}). (b) The temporal evolution of the $B_z$ field for the collisionless (blue) and collisional (green) cases. (c) The temperatures as functions of time for the collisionless (blue) and collisional (green) plasmas, as well as the solution obtained by solving the isotropization equation.}
\label{fig-weibel-collisions}
\end{figure}

Figure \ref{fig-weibel-collisions} presents an example of a 1D PIC simulation of Weibel instability. The simulation considers a stationary plasma with a uniform density of $10^{18}~\rm cm^{-3}$, anisotropic temperatures of $T_y=100$ eV and $T_x=10$ eV, and an immobile ion background. The $B_z$ field is resolved along the $x$ direction with a cell size of $1/64~c/\omega_p$. The $k$-resolved growth rate of the instability is plotted in Fig. \ref{fig-weibel-collisions}(a), which agrees with the prediction of kinetic theory (Eqn. \ref{eqn-disper-weibel}). The temperature anisotropy in this case is $A=9$, which is much smaller than the $A\approx140$ observed in the CFI example shown in Fig. \ref{fig-CFI}, resulting in a narrower unstable region and a much reduced growth rate.

It should be noted that the examples presented thus far do not consider the effects of collisions. In general, collisions tend to reduce the growth rate and generate larger filaments by moving the most unstable mode toward smaller $k$ values. To examine the impact of collisions on Weibel instability, we conducted a PIC simulation with the same parameters as before but included binary collisions \cite{nanbu_theory_1997,sentoku_numerical_2008}. The modified growth rate extracted from this simulation is shown in Fig. \ref{fig-weibel-collisions} for comparison, where both the maximum growth rate and the wavenumber of the most unstable mode are reduced by approximately a factor of two. Fig. \ref{fig-weibel-collisions}(b) displays the temporal evolution of the $B_{z,\rm rms}$ fields in both cases, revealing that the magnetic field is amplified by over two orders of magnitude in the collisionless plasma, while the amplification is suppressed in the collisional plasma. Fig. \ref{fig-weibel-collisions}(c) shows the isotropization of these plasmas. In the collisionless plasma, the temperature anisotropy remains unchanged for the first $\sim$10 ps when the magnetic field is weak, and encounters a sudden drop when the instability approaches saturation. Thereafter, the anisotropy drops more slowly. The temperature anisotropy drops faster in the collisional plasma, mainly due to collisions. For comparison, we also plotted the temperatures calculated using the isotropization equation in the NRL plasma formulary \cite{richardson_2019_nodate}.

\section{Hierarchy of kinetic instabilities in tunnel-ionized plasmas}\label{sec4-hierarchy}
Previous theoretical work has focused on investigating kinetic instabilities that arise in single-photon photoionized plasmas, such as those ionized by an intense X-ray pulse generated by a free-electron laser \cite{bychenkov_distinctive_2006,andriyash_evolution_2008}. These studies often assume an initial electron velocity distribution based on the single-photon ionization theory (see Sec. \ref{sec-single-photon}), after which the growth rate of the kinetic instabilities is analyzed assuming collisions are not important; in other word the instabilities grow and saturate in less than $\nu_{ee}^{-1}$ where $\nu_{ee}$ is the electron-electron collision frequency. Here by using PIC simulations, we self-consistently model both the strong-field tunnel ionization process and the subsequent evolution of the electron velocity distribution under the influence of instabilities. In this section, we demonstrate that all three kinetic instabilities can grow in the highly non-thermal and anisotropic photoionized plasma, through both self-consistent PIC simulations and experiments. Moreover, the different prerequisites and growth rates of these instabilities lead to a hierarchical structure where the TSI and CFI grow first, followed by the thermal WI.

\subsection{Onset of streaming and current filamentation instability in photoionized plasmas}\label{sec4-onset}
\begin{figure}[h]%
\centering
\includegraphics[width=0.9\textwidth]{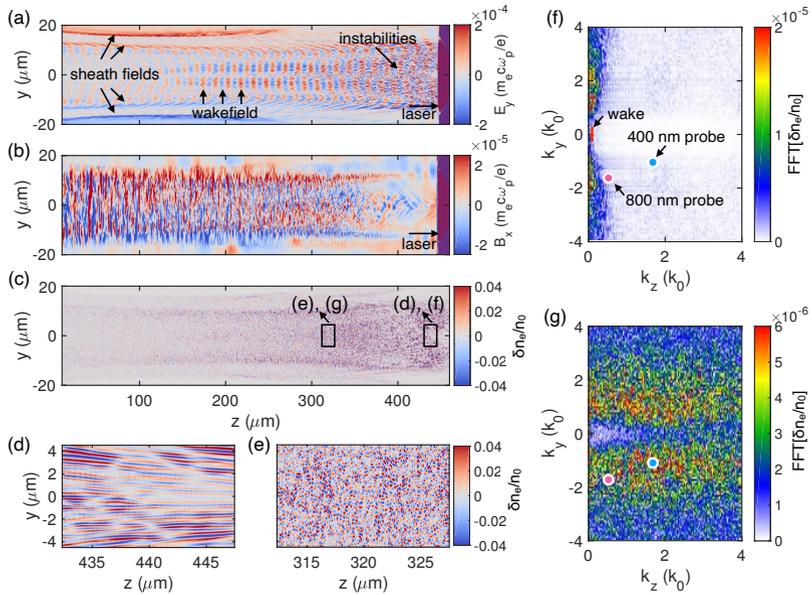}
\caption[Self-consistent PIC simulation of onset of TS and CFI in photoionized plasmas.]{
Self-consistent PIC simulation of the onset of two-stream and current filamentation instabilities in photoionized plasmas. (a)-(b) Electric and magnetic fields in the photoionized plasmas induced by a circularly polarized laser pulse propagating towards the right, with the laser pulse almost exiting the simulation box. The wakefields driven by the laser and the quasi-static sheath fields are highlighted in (a). (c) Density fluctuations associated with the instabilities. The regions marked by the boxes in (c) are magnified in (d) and (e), with the corresponding 2D k-space presented in (f) and (g), respectively at time $t=50$ fs and 400 fs after the peak of the ionization producing pulse has passed through the probed region. By 400 fs, the wake centered at $k_y=0$ excited by the pump pulse has already become broad in $k_z$ space in the presence of the TSI and CFI modes. The probed $k$'s in the Thomson scattering measurements are indicated in (f) and (g). 
}
\label{fig-PIC-instability}
\end{figure}

Figure \ref{fig-PIC-instability} displays the results of a 2D particle-in-cell (PIC) simulation investigating the kinetic instabilities following photoionization, in the strong field, long wavelength (tunneling) limit (discussed in Sec. \ref{sec-tunneling}) where helium gas subjected to a circularly polarized, 0.8 $\rm\mu m$ laser pulse intense enough to ionize both He electrons during the rising intensity of the pulse \cite{zhang_ultrafast_2019}.

The initial distribution function of the plasma is like that depicted in Fig. \ref{fig-an-anisotropicEVD}(b), which consists of radially counterpropagating streams, triggering rapid growth of both TSI and CFI. In this 2D setup, the counterpropagating electron streams are predominantly oriented along the $y$ direction, with the streaming (filamentation) instability exhibiting a wave vector primarily parallel (perpendicular) to the streaming direction. In the linear regime, these instabilities grow independently and the TSI produces fluctuating density strips along the $z$ axis [see Fig. \ref{fig-PIC-instability} (a) and (d)]. However, as they progress into the nonlinear regime, they may couple via the $\bf J\times B$ force and produce small-scale, randomly distributed speckles of density fluctuations in the $y$-$z$ plane, as seen in Fig. \ref{fig-PIC-instability}(e). From Fig. \ref{fig-PIC-instability}(a) one can see that the streaming instability quickly saturates and damps [in about 1 ps or 300 $\rm\mu m$ in space, due to the relaxation of the counterpropagating streams caused by collisionless phase space diffusion caused by electron trapping as shown in Fig. \ref{fig-two-stream} (d). We anticipate a similar temporal behavior for the filamentation instability, which is also driven by these streams. However, the magnetic field $B_x$ continues to grow beyond 1 ps, as illustrated in Fig. \ref{fig-PIC-instability}(b), suggesting that the Weibel instability, driven by a decreased but finite temperature anisotropy of the electrons, becomes the dominant instability in the plasma at later stages. In Fig. \ref{fig-PIC-instability} (d) and (e), the density fluctuations are shown in detail, with corresponding 2D $k$-space shown in Fig. \ref{fig-PIC-instability} (f) and (g) to highlight the transition of unstable region. In the next section we will show that the density fluctuations associated with these instabilities can be measured using Thomson scattering of an externally synchronized probe. The wave vectors being probed at oblique angles with various $k_y$ and $k_z$ values at locations designated by the circles in Fig. \ref{fig-PIC-instability}(f) and (g). These measurements allowed us to concurrently probe the TSI and CFI, despite selecting a particular scattering angle to fix the probed wave vector $k$, as these instabilities differ in their frequencies as discussed in Secs \ref{sec3-streaming} and \ref{sec3-CFI}.

\subsection{Thomson scattering measurements of streaming and current filamentation instabilities}\label{sec4-measurement}
In this section, we present the ultrafast measurement of the kinetic instabilities in photoionized plasmas using Thomson scattering \cite{zhang_ultrafast_2019}. By utilizing a probe pulse with a duration of $\lesssim100$ fs, which was split from the same laser system as the pump laser that produces the plasma, we can precisely control the timing jitter between the two pulses down to few fs level. The short probe duration as well as the precision synchronization allow us to capture the first measurement of these kinetic instabilities on a timescale that is fast enough to measure their growth rate and unstable frequency spectrum during the growth phase in relatively dense plasmas ($10^{18}~\rm cm^{-3}$).

\begin{figure}[h]%
\centering
\includegraphics[width=1.0\textwidth]{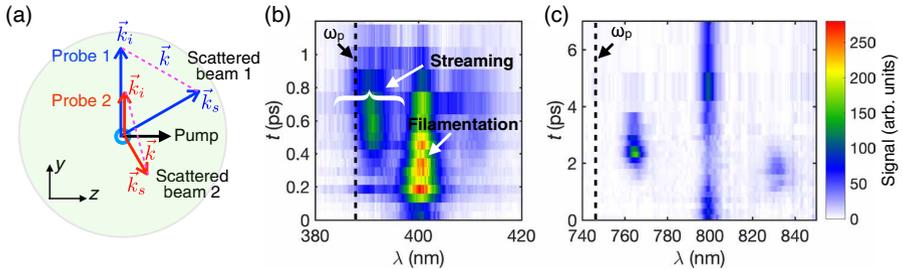}
\caption[TS diagram and examples of measured TS spectra.]{Thomson scattering (TS) diagram and representative TS spectra for helium plasma. (a) $\bf k$-matching diagram showing the experimental setup, where a helium plasma generated by a 50-fs, 800-nm circularly polarized (CP) or linearly polarized (LP) pump laser is probed using a 400-nm (probe 1) or 800-nm (probe 2) laser, respectively. The TS spectra obtained with a variable delay between the pump and probe lasers are shown for (b) CP pump and (c) LP pump. The dashed lines indicate the expected plasma frequency corresponding to the plasma density. The entire dataset is obtained by averaging 20 individual scattering events with timing steps of 50 to 200 fs. Time t = 0 is defined as the time when pump and probe overlap at the ionization front, as seen in a shadowgram formed by the probe beam at the same location.}
\label{fig-ts-setup}
\end{figure}

The experimental directions of the incident laser beams and the collected scattered light photons are shown in Fig. \ref{fig-ts-setup}(a), where two $k$-matching diagrams are presented for the two probe lasers used, 0.4-$\mu$m (probe1) and 0.8-$\mu$m (probe2), respectively. The probe bandwidth is approximately 5 nm, and the pulse width is nominally 100 fs. Scattered light is collected at $60^\circ$ (scattered beam 1) for the CP case and $150^\circ$ (scattered beam 2) for the LP case, thereby measuring waves with vectors satisfying the $k$-matching condition $k_s=k_i\pm k$. The probed wave vectors are indicated by the magenta dashed lines, and their locations in the 2D $k$ space are marked by circles in Fig. \ref{fig-PIC-instability}(f) and (g). 

There is another point worth noting. In this experiment the pump pulse is intense enough to not only ionize both electrons of He but it is also short enough to excite a quasi-linear wake whose normalized amplitude $n_1/n_0$ scales as $a_0^2$ which for our $a_0$ of 0.2 is $4\%$. The wake is the fastest excited mode growing on a timescale of $\omega_p^{-1}$ or $\sim10$ fs. In a collisionless plasma, linear wake can last for a long time since they do not suffer Landau damping owing to its relativistic phase velocity \cite{zhang_femtosecond_2017,zhang_evolution_2018}. But what is interesting is that the kinetic instability modes occurring as self-organized structures appear almost as fast and spread through the $k$ space as can be seen from simulation results shown in Fig. \ref{fig-PIC-instability} and confirmed by the temporal evolution of the frequency spectra obtained by Thomson scattering shown in Fig. \ref{fig-ts-setup}. This fact had not been recognized before by the plasma accelerator community to our knowledge when the laser pulse is used to both create the plasma and excite linear wakes ($n_1/n_0\ll1$).

\begin{figure}[h]%
\centering
\includegraphics[width=0.95\textwidth]{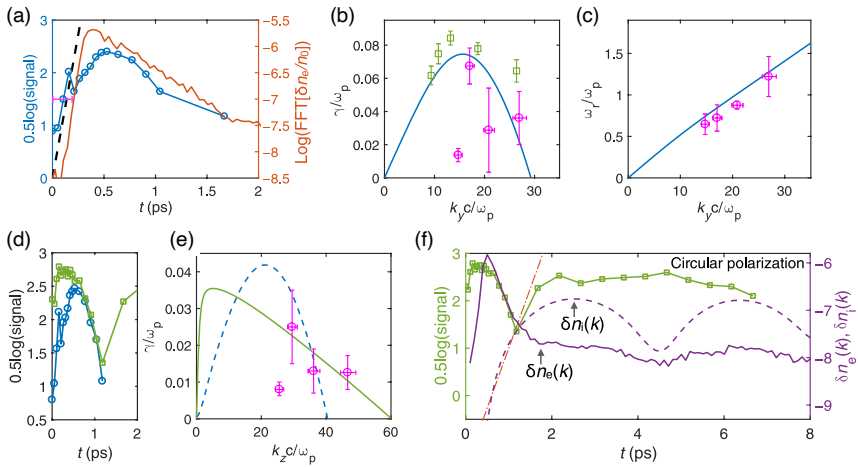}
\caption[Thomson scattering measurements of kinetic instabilities.]{Thomson scattering measurements of kinetic instabilities. (a) Temporal evolution of the magnitude of the electron feature (blue circles) compared with the density fluctuation magnitude obtained from a 2D PIC simulation (red line) and the exponential growth of a linear wave from kinetic theory (black dashed line) for the same probed value of $k$. The horizontal error bars denote the uncertainty in determining the completion time of ionization. (b) Measured (magenta circles), predicted (blue line), and simulated (green squares) growth rates of the instability, with horizontal error bars indicating the uncertainty of density measurement and vertical error bars representing the $\pm\sigma$ confidence interval of the deduced growth rate. The blue dashed line shows the growth rate of the zero-frequency mode of the streaming instability. (c) Measured (magenta circles) and predicted (blue line) frequencies of the streaming instability for the same range of densities as in (b). (d) Measured magnitude of the zero-frequency feature (green) and the electron feature (blue) within the first 2 ps. (e) Initial growth rates of the filamentation instability obtained from the measurements (circles) and the calculated solution of the dispersion relation (green line). The blue dashed line represents the growth rate of the non-oscillating mode of the streaming instability. (f) Measured magnitude of the zero-frequency feature as a function of time (green squares) compared with the evolution of the amplitude of the electron density fluctuation (solid purple line) and the ion density fluctuation (dashed purple line) in the simulation at the same $k$ that is being probed in the experiment. The red dotted-dashed line shows the maximum growth rate of the Weibel instability calculated using the simulated EVD at $t$=1 ps.}
\label{fig-validate-theory}
\end{figure}

Figure \ref{fig-validate-theory}(a) presents the blue-shifted electron satellite amplitude as a function of the probe delay in a logarithmic scale (blue circles). The red line represents the wave amplitude obtained from a 2D PIC simulation performed with the same experimental parameters (Fig. \ref{fig-PIC-instability}). The black dashed line shows the initial growth rate predicted by the kinetic theory. Apart from a 100-fs offset at $t=0$, the temporal variation of the measured signal agrees well with the kinetic theory and simulations. The instability grows at the expense of the electrons’ directional energy, so its growth rate decreases until it saturates at $\sim0.5$ ps ($\sim70~\omega_p^{-1}$) and then starts to damp. The nonlinear phase follows, during which the distribution function changes significantly due to wave-particle interactions, lasting for another 0.5 ps. Some of the electrical field energy of the unstable waves is returned to the electrons via electron trapping, resulting in transverse phase space diffusion leading to the onset of isotropization.

We conducted measurements at four different plasma densities and determined the initial growth rate and the frequency of the fastest growing mode at the measured $k$ of the streaming instability for each density. The results are presented as magenta circles in Fig. \ref{fig-validate-theory} (b) and (c), respectively. The blue lines represent the predictions of the kinetic theory, while the green squares in Fig. \ref{fig-validate-theory}(b) depict the 2D simulation results of the growth rate of the instability for the same $k$ as measured in the experiment. Here, $k_0 = 2\pi/\lambda_0$ is the wave vector of the 800-nm laser. The frequency of the instability shown in Fig. \ref{fig-validate-theory}(c) was calculated using the spectral shift of the peak of the blue satellite before the instability saturates [see Fig. \ref{fig-validate-theory}(b)]. The measured initial growth rates show reasonable agreement with the theory. The differences may be partially caused by the fact that collisions may have reduced the growth rate of the instability, which may explain why the measured growth rate is smaller than the analytical and computational predictions.

We found excellent agreement between the measured and predicted frequencies of the streaming instability. The reason for this lies in the expression for the instability frequency, which is given by $\Delta\omega\approx {\bf k\cdot v_d}=k_yv_d$, where $v_d\approx 0.04c$ is the drift velocity between the co-propagating $\rm He^{1+}$ and $\rm He^{2+}$ electrons, primarily determined by the ionization process itself. Due to the significant spread in $v_d$, a broad range of unstable streaming modes is observed to grow (at a given $k$) in the TS electron spectrum [Fig. \ref{fig-PIC-instability}(b)]. Notably, the frequency of the streaming modes induced by the counter-streaming $\rm He^{1+}$ ($\rm He^{2+}$) species is zero, as the two beams are symmetric and hence do not contribute to the measured electron feature.

The phase velocity of the TSI is given by $v_\phi\equiv\omega/k_y\approx v_d$, where $v_d$ is the drift velocity between the copropagating $\rm He^{1+}$ and $\rm He^{2+}$ electrons. This suggests a strong resonance between the copropagating $\rm He^{1+}$ and $\rm He^{2+}$ electrons. This resonant interaction causes the initially double donut-shaped electron distribution to undergo collisionless phase space diffusion, resulting in a single quasi-Maxwellian distribution within approximately 1 ps.

We also investigated the CFI initiated by the radially counterpropagating streams. The measured growth of the zero-frequency feature (green squares) and the electron feature (blue circles) are shown in Fig. \ref{fig-validate-theory}(d). For the $k$ values being probed, the zero-frequency filamentation mode is the first to appear above the measurement threshold, followed by the streaming mode, both of which have similar growth rates. These modes exhibit rapid growth and then decay within approximately 1 ps, which is shorter than the ion plasma period ($2\pi\omega_{pi}^{-1}\approx3$ ps). These observations suggest that the zero-frequency mode corresponds to the filamentation instability, rather than the usual ion acoustic waves.

In Fig. \ref{fig-validate-theory}(e), the initial growth rate of the filamentation mode is presented, where the magenta circles indicate the measurements, and the green line shows the solution of the dispersion relation. The agreement between the measurements and the kinetic theory is reasonable. It is worth noting that the kinetic theory predicts the possibility of a non-oscillating branch of the streaming instability. The growth rates of both branches are very similar; hence, this non-oscillating branch is also likely to contribute to the measured zero-frequency feature. However, the recurrence of the zero-frequency feature in the measurements, as shown in Fig. \ref{fig-validate-theory}(f) by the green curve, is the strongest evidence of the CFI and WI, as the recurrence of the streaming mode is not expected once the counterpropagating streams no longer exist.

The purple lines in Fig. \ref{fig-validate-theory}(f) represent the amplitudes of the density fluctuations of electrons ($\delta n_e$) and ions ($\delta n_i$) extracted from a 2D PIC simulation that models the experiment (without ionization but includes mobile ions), at the same $k$ as that being probed in the experiment as a function of time. The evolution of electron density fluctuation $\delta n_e(k_{\rm expt})$ shows a peak at $t\approx0.5$ ps, which well tracks the first peak of the measured zero-frequency feature. This non-oscillating electron density fluctuation is due to the electrostatic component of CFI, which arises from the asymmetric counterpropagating $\rm He^{1+}$ and $\rm He^{2+}$ electron streams. As time progresses, the scale length of these density fluctuations evolves to larger $k$, allowing the Weibel instability to grow to detectable levels through the ion density fluctuations. In the experiment, this manifests as the recurrence of the zero-frequency feature, while in the simulations, this is seen as the growth of the ion density fluctuations after the first $\sim$2 ps. The theoretical growth rate of the Weibel instability calculated using the EVD at 1 ps is shown as a red dotted-dashed line. While the unstable range of the Weibel instability covers $0<k<\sqrt{A}~\omega_p/c$, which is much smaller than the probed $k$- it has been shown that the unstable $k$ of the Weibel instability can extend to larger $k$ through turbulence cascade \cite{mondal_direct_2012}.

\subsection{Relaxation of the tunnel-ionized plasma electrons}\label{sec4-relaxation}
The growth of TSI generates electric fields, which can scatter electrons resulting in a relaxation of the original non-thermal distribution towards a Maxwellian distribution. Moreover, the temperature anisotropy drops due to the growth of CFI and WI, which self-generate magnetic fields, further impacting the electron dynamics. We have tracked the evolution of the electron velocity distributions and temperature anisotropy in the 2D simulation. The simulation excluded Coulomb collisions to isolate the effect of the instabilities on the temperature anisotropy.

\begin{figure}[h]%
\centering
\includegraphics[width=0.95\textwidth]{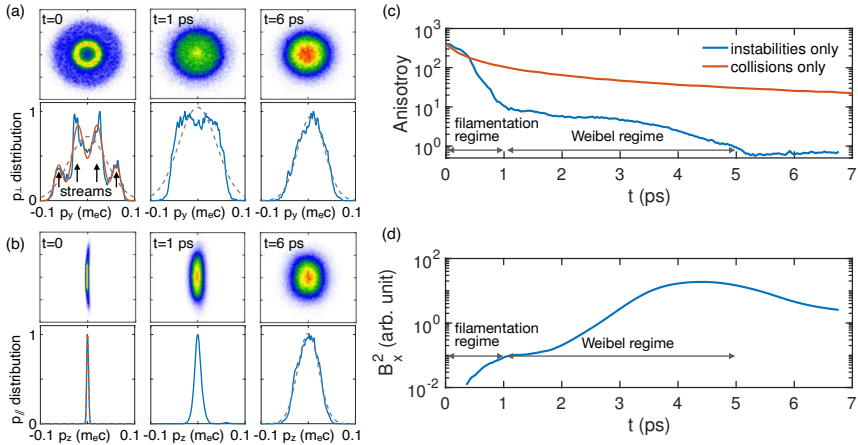}
\caption[Thermalization and isotropization of photoionized plasmas.]{Thermalization and Isotropization of Photoionized Plasmas. (a) Upper panel: Snapshots of transverse momentum space at representative times. Lower panel: Projected distribution $f(p_y)$. (b) Upper panel: Snapshots of $p_x{\text -}p_z$. Lower panel: Projected distribution $f(p_z)$. (c) Temperature anisotropy evolution deduced from instabilities only (blue) and collisions only (red) simulations. (d) Growth of self-generated magnetic fields due to filamentation and Weibel instabilities.}
\label{fig-isotropization}
\end{figure}

The results of the simulation are presented in Fig. \ref{fig-isotropization}. The upper row of Fig. \ref{fig-isotropization}(a) shows snapshots of the transverse momentum space at $t$=0, 1, and 6 ps, while the lower row shows the corresponding projected distribution $f(p_y)$ within a representative slab of the plasma. Initially, the $p_y$ distribution exhibits a highly nonthermal shape that can be approximated by four drifting Maxwellian beams in the transverse plane, deviating significantly from the Maxwellian distribution shown by the gray dashed line [see also Fig. \ref{fig-an-anisotropicEVD}(b)]. However, within only a few ps, the collisionless phase space diffusion, caused by Landau damping and particle trapping [see Fig. \ref{fig-two-stream}(d)] by the waves excited by the streams, leads to the multiple beam structure disappearing, and the distribution approaching a quasi-Maxwellian one. Fig. \ref{fig-isotropization}(b) shows the corresponding $p_z$ distribution. The plasma has an extremely low temperature in the laser propagation direction (indicated by the red dashed line). The growth of CFI/WI magnetic fields and the subsequent $\bf v\times B$ motion of electrons cause the temperature $T_\perp$ to drop and $T_\parallel$ to increase significantly.

The simulation results demonstrate that kinetic instabilities play a significant role in the rapid decrease of plasma anisotropy from an initial value of a few hundreds to approximately 10 within 1 ps, as illustrated by the blue line in Fig. \ref{fig-isotropization}(b). Although the WI continues to further isotropize the plasma, it does so at a reduced rate, ultimately reducing the anisotropy to less than 1 in approximately 7 ps. The magnetic fields, as shown in Fig. \ref{fig-isotropization}(c), exhibit two distinct growth phases corresponding to the CFI and WI.

To confirm that the observed rapid isotropization is solely due to kinetic instabilities, we conducted an additional simulation using a preionized plasma with a comparable initial EVD, as represented by the red lines in Fig. \ref{fig-isotropization}(a). This simulation included only Coulomb collisions (both $e{\text -}e$ and $e{\text -}i$), and the resulting anisotropy is shown by the red line in Fig. \ref{fig-isotropization}(c). In the absence of kinetic instabilities, electrons exchange energy and momentum solely through collisions, causing them to isotropize after several tens of picoseconds. The comparison between the two simulations indicates that over the range of plasma densities used, collisions do not play a significant role during the first ten picoseconds after plasma formation, and isotropization is primarily driven by kinetic instabilities during this time. However, collisions will eventually thermalize the plasma after the saturation of the instabilities.

\begin{figure}[h]%
\centering
\includegraphics[width=0.95\textwidth]{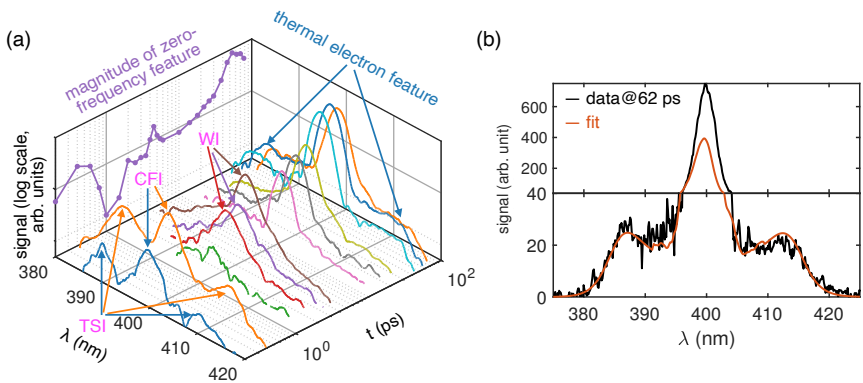}
\caption[Long-term evolution of Thomson scattering spectra of a plasma ionized by a CP laser.]{Long-term evolution of Thomson scattering spectra of a plasma ionized by a CP laser. (a) Thomson scattering spectra collected from a helium plasma with $n_e\approx6\times10^{18}~\rm cm^{-3}$. Each line represents the average of 10 shots at the same delay. Note that 12 out of the total 27 lines of the dataset are shown to improve the clarity. The purple curve in the top left projection plane shows the magnitude of the zero-frequency feature. (b) Fitting the electron features of TS spectra at a late delay indicated by the red line in (a). The black line represents the measured spectra and the red line is the best fit using a Maxwellian distribution.}
\label{fig-long-term}
\end{figure}

The long-term thermalization of photoionized helium plasma produced by circularly polarized pump pulses was also investigated in the experiment. Figure \ref{fig-long-term} (a) displays the evolution of Thomson scattering spectra, which is the same dataset as in Fig. \ref{fig-ts-setup}(b) but plotted on a log scale. As previously discussed, the simultaneous growth of the electron and zero-frequency features within the first ps correspond to the TSI and CFI, respectively. The second peak of the zero-frequency feature appears due to the WI and lasts for about 10 ps, which agrees with the simulation result shown in Fig. \ref{fig-isotropization}(d).

Interestingly, an electron feature with distinct peaks at the expected Bohm-Gross frequency and a much more enhanced ion feature appeared in the Thomson scattering spectra after about 40 ps. This indicates that the plasma had become thermalized at this time. Fig. \ref{fig-long-term}(b) shows an example of the measured TS spectra at a late delay (62 ps) and the corresponding best fit. Analysis of spectra at different densities consistently shows the expected thermalized temperature of 80-100 eV, with a fixed amount ($\sim35\%$) plasma density drop compared to the initial density right after ionization. Notably, also we performed the same measurement for plasma ionized by a linearly polarized laser, but no electron spectral feature characteristic of a Maxwellian plasma was observed until the limit of the measurement of $\sim$97 ps. This implies that it takes a much longer time for OFI plasmas in the LP case to become thermalized or unmagnetized, despite its initial distribution being closer to a Maxwellian one.

\section{Ultrafast self-organization of magnetic fields in photoionized plasmas}\label{sec5-self-organization}
In the previous section, we have shown that photoionized plasmas exhibit a hierarchy of kinetic instabilities, due to the initial highly nonthermal and anisotropic distributions of photoionized plasma electrons. These instabilities lead to rapid thermalization and isotropization of the plasma in a few ps. However, the remnant temperature anisotropy, $A<2$, is still able to excite the thermal Weibel instability, which generates magnetic fields in the plasma that persist for a much longer time. The formation of these magnetic fields, which can exhibit a well-defined topology, is another example of the self-organization of photoionized plasmas. In this section, we delve deeper into this topic, examining the physical mechanisms underlying this self-organization and the impact it has on the overall evolution of the plasma.

\subsection{Ultrafast probing of CFI/Weibel magnetic fields}\label{sec5-probing}
Although extensively studied in theory and particle-in-cell simulations, experimental study of the thermal WI has been extremely limited. In the last decade, several experimental approaches have been established, with most of them being suitable for the study of either ion or electron CFI. 

\begin{figure}[h]%
\centering
\includegraphics[width=0.95\textwidth]{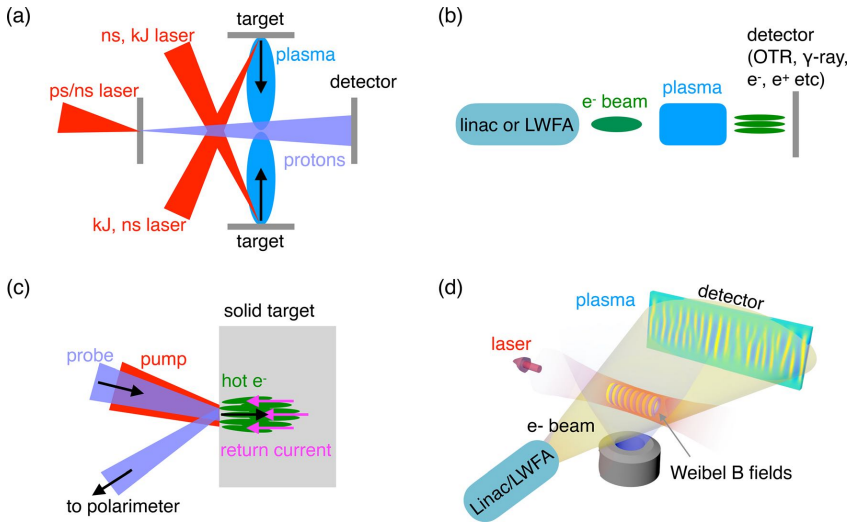}
\caption[Overview of experimental approaches for studying CFI/Weibel instability and our new platform.]{Overview of experimental approaches for studying CFI/Weibel instability and our new platform. (a) Colliding plasma approach for probing ion CFI using laser-driven proton radiography or Thomson scattering. (b) Relativistic beam-plasma interaction for studying relativistic electron CFI. (c) Laser-solid interaction for investigating non- to quasi-relativistic electron CFI. (d) Our new platform that enables the investigation of thermal electron Weibel instability.}
\label{fig-probing}
\end{figure}

The method presented in Fig. \ref{fig-probing}(a) utilizes multiple laser pulses with a total energy from kJ to MJ to create ablation plasmas by blasting two solid targets, such as CH foils, arranged face-to-face and separated by a few millimeters to centimeters. The resulting plasmas expand and interpenetrate through each other, triggering the growth of ion CFI since the energy in these flows is mainly carried by ions. The growth of magnetic fields is then probed by proton bunches generated by a separate synchronized laser pulse that accelerates protons from a thin contamination layer of hydrocarbons on the rear surface of a secondary target that is place in the vicinity of but orthogonal to the plane that contained the colliding plasmas. The protons in the tens of MeV range traverse the collision region where they are deflected by electric and magnetic fields producing density striations from which one can determine the magnitude of the deflecting fields. This diagnostic is called laser-driven proton radiography \cite{fox_filamentation_2013}. The characteristic filamentary structure of CFI-induced magnetic fields and the Biermann battery effect (magnetic fields due to the  $\nabla n_e\times\nabla T_e$ source term \cite{haines_magnetic-field_1986}) have been identified using this platform \cite{huntington_observation_2015}. Recently, a modified version of this platform was employed where the two solid targets were tilted to produce plasma flows that collided at a $130^\circ$ angle. Thomson scattering of an external optical probe was used to record the evolution of current filaments moving through a fixed scattering volume \cite{swadling_measurement_2020}. These approaches and their variations have been effective in probing high energy density plasmas. However, their limited adoption is mainly due to the requirement for energetic (kJ-MJ) lasers, which are only available at large facilities and typically operate at low repetition rates (on the order of a few shots per day).

The second approach, depicted in Fig. \ref{fig-probing}(b), offers a promising avenue for studying relativistic electron CFI. In this method, a relativistic electron bunch from either a linear radio-frequency (RF) or laser wakefield accelerator is sent through a stationary plasma (either underdense $n_b>n_p$ or overdense $n_b<n_p$) \cite{allen_experimental_2012,huntington_current_2011}. The electron bunch undergoes modulation via a streaming instability and eventually breaks into filaments due to electron CFI. In the underdense case, the filamented beam passes through a metallic foil, generating optical transition radiation (OTR) that is used to visualize the break-up of the beam after its interaction with the plasma \cite{allen_experimental_2012}. In the overdense case, an ultra-relativistic electron bunch is focused onto a metallic foil where it generates high energy x-rays that are interpreted as arising from electron CFI which generates up to $10^4$ Tesla magnetic fields that bend the trajectories of beam electrons emitting high energy synchrotron radiation \cite{benedetti_giant_2018,sampath_extremely_2021}.

The third approach, illustrated in Fig. \ref{fig-probing}(c), enables the study of electron CFI in the quasi-relativistic regime. This technique utilizes an ultra-intense laser pulse ($a_0>1$) that is focused onto a solid target to drive hot electrons with energy ranging from a few tens keV to a few MeV in the forward direction at the critical density surface of the target. These forward propagating electrons and the induced return current of the background cold electrons trigger the growth of both TSI and CFI. The turbulent magnetic fields on the surface of the target can be measured using optical polarimetry. The magnetic field structures attributed to the CFI and Weibel within the target are far harder to measure. Therefore, this potentially promising platform has largely been explored through PIC simulations \cite{mondal_direct_2012}.

Despite these very clever techniques, conclusive demonstration of the thermal WI driven by a temperature anisotropy in a stationary plasma as originally envisioned by Weibel has been elusive until recently, because experimental tools to first initialize known anisotropic electron temperature distributions and to measure the growth, saturation, and decay of the magnetic field structures with ultrafast time resolution were not available. Now we present a recently developed diagnostic technique, sketched in Fig. \ref{fig-probing}(d), for unambiguous verification of the thermal electron Weibel instability. We have already discussed how a stationary plasma with controllable EVD can be produced using high-field photoionization, now we show how the magnetic fields generated in the plasma can be recorded by measuring the deflections of high-quality relativistic electrons from a linear accelerator with $\rm\mu m$ and ps spatiotemporal resolution \cite{zhang_measurements_2020,zhang_mapping_2022}.

\subsection{Formation of magnetic helicoids in plasmas ionized by circularly polarized lasers}\label{sec5-CP}
Using the experimental setup depicted in Fig. \ref{fig-probing}(d), we have recently conducted a series of experiments that have for the first time isolated the electron thermal WI \cite{zhang_measurements_2020}, which supported the formation of quasi-static structures exhibiting a topology consistent with predictions of kinetic theory \cite{romanov_self-organization_2004}. The first experiment involved the photoionization of helium gas using a circularly polarized Ti:Sapphire laser pulse, with 50 fs pulse length, 22 $\rm\mu m$ spot size ($w_0$) and a peak intensity of $\sim2.5\times10^{17}~\rm W/cm^2$. This laser intensity was sufficient to rapidly ionize both electrons of helium atoms through tunneling ionization during the pulse rise time. The plasma density was adjusted within the range of (0.3-1.5)$\times10^{19}\rm cm^{-3}$ by varying the backing pressure of the supersonic nozzle. To measure the fields within or on the edge of the photoionized plasma, we employed an ultrashort electron bunch with a peak energy of 45 MeV, a pulse duration of 1.8 ps, a total charge of $\sim$30 pC, and negligible energy spread ($0.5\%$), delivered by a RF linear accelerator. The peak current of the bunch was only 17 A, which was insufficient to excite a measurable relativistic plasma wave in via a self-modulation instability \cite{fang_seeding_2014}. The probe was orthogonally incident on the plasma, and the resulting deflection of the probe electrons by the electromagnetic fields inside the plasma was captured. However, the relatively long probe was insensitive to fast-oscillating electric fields caused by wakes (the probing of which requires femtosecond electron bunches \cite{zhang_femtosecond_2017}), as its length was much greater than the oscillation period of the fields ($2\pi\omega_p^{-1}$), leading to the cancellation of deflections of different longitudinal slices of the probe and a net deflection approaching zero \cite{zhang_capturing_2016}. Therefore, the electron probe was only effective in capturing quasi-static magnetic fields inside the plasma and sheath electric fields on the edge of the plasma.

\begin{figure}[h]%
\centering
\includegraphics[width=0.95\textwidth]{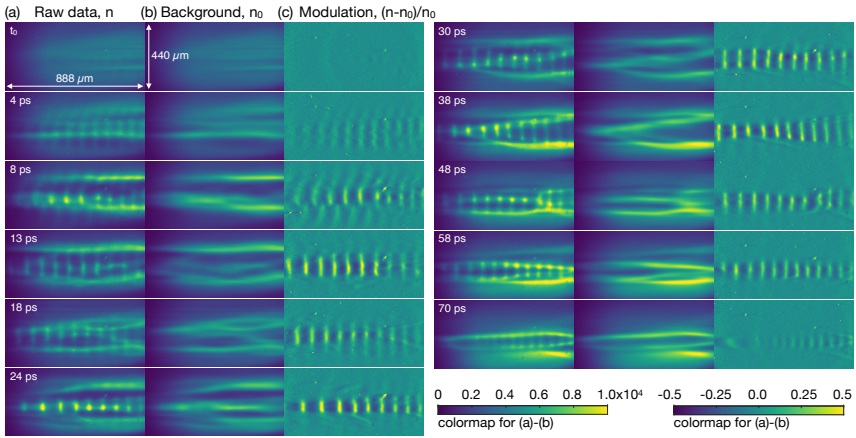}
\caption[Snapshots of Weibel magnetic fields.]{Time-resolved snapshots of flux modulations induced on the probe electron beam by Weibel magnetic fields in helium plasma, ionized by circularly polarized, 0.8-$\rm\mu m$ laser pulses. The snapshots were taken at various time intervals, as indicated in the figure, while the laser propagated from right to left. The induced electron deflections are clearly visible in the snapshots.}
\label{fig-weibel-cp-data}
\end{figure}

The temporal evolution of the electron probe bunch after traversing the plasma was captured by varying the delay of the probe with respect to the ionizing laser. The resulting time-resolved snapshots are presented in Fig. \ref{fig-weibel-cp-data}. Figure \ref{fig-weibel-cp-data}(a) displays the raw data, which corresponds to the modulated flux of the electron probe due to the fields in the plasma. At $t_0$, defined as the moment when the probe overlaps the laser at the interaction point, only a few faint horizontal strips are observed. These may arise due to deflection caused by the sheath electric field. Later, these faint structures, referred to as large-scale structures, become more evident. Additionally, small-scale structures such as vertically aligned strips emerge on top of the large-scale structures and continue growing until they reach saturation at around $\sim20$ ps. Since the large- and small-scale structures have orthogonal orientations, it is possible to distinguish between them by smoothing the raw image line by line along the horizontal direction. The resulting images are presented in Fig. \ref{fig-weibel-cp-data}(b) and (c) for the large- and small-scale structures, respectively. The growth of Weibel magnetic fields with wave vectors parallel to the laser axis is favored by the initial EVD of the photoionized plasma. Therefore, we interpret the small-scale structures as arising from the Weibel magnetic fields, while the large-scale structures can be considered an effective background and its origin will be discussed later.

The relative modulation of the flux of the electron probe, $(n-n_0)/n_0$, presented in Fig. \ref{fig-weibel-cp-data}(c), reveals the projected structure of the magnetic fields and is proportional to the field strength. The data in Fig. \ref{fig-weibel-cp-data}(c) presents a clear pattern in which the flux modulation magnitude of the probe beam increases from a detection threshold level ($\approx0.03$) to reach its peak at $\approx24$ ps, slowly decreasing in tens of ps afterwards. The average wavelength of the magnetic fields for this dataset is $\approx80~\rm\mu m$, which implies that $k\approx 0.18~\omega_p/c$ for $n_e\approx5\times10^{18}~\rm cm^{-3}$. This relation holds for all the five datasets measured within the density range (0.3-1.5)$\times10^{19}~\rm cm^{-3}$. For a collisionless plasma, the most unstable mode is $k_m=\sqrt{A/3}~\omega_p/c$. This suggests that the plasma temperature anisotropy has dropped to $A\approx0.1$ by the time the magnetic fields become detectable if we assume $k\sim k_m$. This rapid drop is attributed to the precursor TSI and CFI with the assistance of collisions.

\begin{figure}[h]%
\centering
\includegraphics[width=0.95\textwidth]{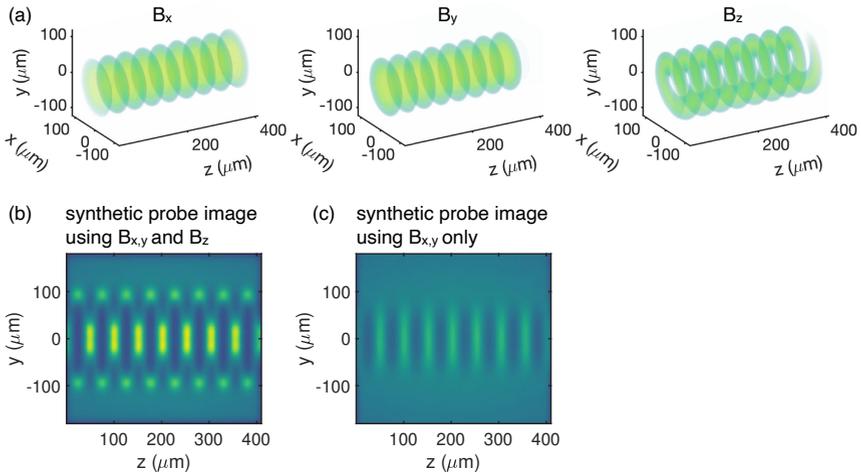}
\caption[Illustration of magnetic helicoids in plasmas.]{Illustration of magnetic helicoids in plasmas. (a) Magnetic field topology consistent with experimental observation, showing the helicoid al structure of the transverse components and the radial dependence of the longitudinal component. (b) Synthetic probe image using the magnetic field shown in (a), including all three components. (c) Synthetic probe image using only the transverse components. The color scales for (b) and (c) are the same for ease of comparison.}
\label{fig-weibel-cp-helicoid}
\end{figure}

From the data, we inferred that the transverse components of the Weibel magnetic field have a helicoidal topology, given by ${\bf B}\approx{\hat x}B_0\cos kz+{\hat y}B_0\sin kz$. Figure \ref{fig-weibel-cp-data}(c) shows vertical strips along $y$ with approximately sinusoidal variations along $z$ in the probe flux modulation data. This occurs because the $B_y$ component deflects the probe electrons, while the $B_x$ component does not, as vprobe is parallel to $B_x$. Although the $B_x$ component is not directly visible to the diagnostic, we can reasonably assume that it has the same form as $B_y$, but with a $\pi/2$ phase shift due to symmetry considerations for a circularly polarized laser. 

The measured transverse components of the Weibel magnetic field exhibit radial dependence, which implies the existence of a longitudinal component $B_z$ to satisfy $\nabla\cdot{\bf B}=0$. A numerical example is presented in Fig. \ref{fig-weibel-cp-helicoid}. In this example, the magnetic fields are assumed to have a peak strength of 0.02 T, a wavelength of 60 $\rm\mu m$, and a radial dependence of $f(r)=\exp(-r^4/\sigma^4)$ with $\sigma=82~\rm\mu m$ with $\sigma=82~\rm\mu m$ which gives a FWHM of 150 $\mu m$. The 3D distribution of $B_x$, $B_y$, and $B_z$ is visualized in Fig. \ref{fig-weibel-cp-helicoid}(a)-(c), respectively. The $B_x$ and $B_y$ components show as parallel disks whereas the $B_z$ component has a helical structure that properly connects the magnetic field lines to satisfy $\nabla\cdot{\bf B}=0$. In contrast to the transverse components, $B_z$ vanishes on the axis and remains small near the axis where the radial gradient of $B_{x,y}$ is small. This implies that in the central region, the deflection of the probe electrons was primarily due to the transverse components. On the edge of the plasma, deflection due to the $B_z$ field becomes dominant.

The magnetic fields presented in Fig. \ref{fig-weibel-cp-helicoid}(a) closely replicate the observed electron flux modulation. To demonstrate this, we tracked the probe beam through the magnetic fields shown in Fig. \ref{fig-weibel-cp-helicoid}(a) using the experimental setup and generated two synthetic images as shown in Fig. \ref{fig-weibel-cp-helicoid}(b) and (c). Figure \ref{fig-weibel-cp-helicoid}(b) includes all three components of the magnetic field, while in Fig. \ref{fig-weibel-cp-helicoid}(c) only the transverse components are utilized.

When all three components are included in the synthetic image, vertical strips are observed near the $z$ axis, as well as two rows of bright spots on the plasma edge at $y\approx\pm100~\rm\mu m$. This is like the data shown in Fig. \ref{fig-weibel-cp-data}(a), although in the data, continuous bright horizontal stripes were observed on the plasma edge instead of separated bright dots. The difference between the two could be due to several factors. One possibility is that the topological structure of the $B_z$ component may be different due to the irregular shape of the plasma boundary, which could have smoothed out the fine structures. Another possibility is that, in addition to the Weibel magnetic field, there may exist other magnetic fields generated via $\nabla n_e\times\nabla T_e$ (the Biermann battery effect \cite{haines_magnetic-field_1986,gregori_generation_2012}) and the electric sheath field on the plasma edge, which also contribute to the observed large-scale structure. Unfortunately, PIC simulations are not possible over the entire time window of the experiment but one can see in the experimental data of Fig. \ref{fig-weibel-cp-data} that the dots show up before the stripes during the 4-8 ps window which could support the above notion of a second mechanism being at work on the plasma edge. The synthetic image presented in Fig. \ref{fig-weibel-cp-helicoid}(c) is generated by considering only the $B_{x,y}$ components. The resulting structure is like the data shown in Fig. \ref{fig-weibel-cp-data}(c), suggesting that these structures are indeed caused by the transverse components of the Weibel magnetic field.

\subsection{Collisional growth rates and saturation of Weibel instability}\label{sec5-collisions}
\begin{figure}[h]%
\centering
\includegraphics[width=0.6\textwidth]{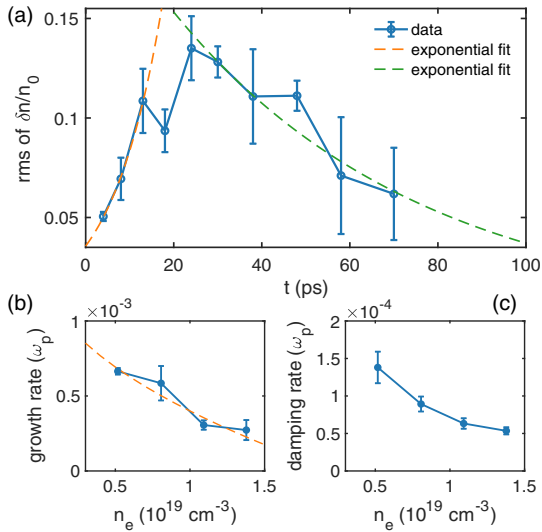}
\caption[Collisional growth and damping rate of Weibel magnetic fields.]{Collisional growth and damping rate of Weibel magnetic fields. Temporal evolution of the measured density modulation (a) and its corresponding growth (b) and damping rate (c). The error bars represent the standard deviation of multiple shots. The dashed line in (b) represents the fit of the growth rate, which takes into account collisions ($\gamma_c=\gamma_0-\nu_e$, see text). The error bars in (b) and (c) represent the standard deviation of the fitting coefficients obtained from (a).}
\label{fig-weibel-cp-g}
\end{figure}

The time-resolved measurement with picosecond resolution enables us to retrieve the growth rate of the WI and compare it with kinetic theory predictions. Since the flux modulation of the relativistic electron probe beam is proportional to the magnetic field strength, we use the deduced flux modulation magnitude to calculate the growth rate of the instability for simplicity. The uncertainty of this treatment is found to be negligible for the parameters used in this measurement, as explained in \cite{zhang_electron_2022}. The temporal evolution of the probe flux modulation magnitude, deduced from Fig. \ref{fig-weibel-cp-data}(c), is plotted in Fig. \ref{fig-weibel-cp-g}(a). The data clearly shows rapid growth, followed by peaking at roughly 20 ps, and then a decay at a smaller rate compared to the growth. The dashed lines in Fig. \ref{fig-weibel-cp-g}(a) represent exponential fits to the data. From these fittings, we extracted both the growth and damping rates of the magnetic fields, which are depicted in Figs. \ref{fig-weibel-cp-g}(b) and (c), respectively.

For a collisionless plasma, the normalized growth rate shown in Fig. \ref{fig-weibel-cp-g}(b) is expected to be constant since it is solely determined by $A$ and $T_\parallel$ (or $T_\perp$), which are determined by the precursor instabilities (e.g., TSI and CFI) arising in the photoionized plasma. By the time the Weibel magnetic field starts dominating the deflection of the probe electrons, the anisotropy level is about the same, as evidenced by the measured constant $k\approx0.18\omega_p/c\sim\sqrt{A/3}\omega_p/c$ for different densities. Previous work \cite{wallace_collisional_1987,mahdavi_collisional_2013,hao_relativistic_2009} and our PIC simulations [see Fig. \ref{fig-weibel-collisions}(a)] show that collisions tend to reduce the growth rate and narrow the width of the unstable spectrum towards smaller $k$. Using the Krook collision model, one can estimate the effective growth rate considering collisions as $\gamma_c=\gamma_0-\nu_e$, where $\gamma_0$ denotes the collisionless growth rate and $\nu_e=(1+Z)\nu_0$ represents the collision rate, which incorporates both the $e{\text -}e$ and $e{\text -}i$ collisions. Here, $\nu_{ee}=\nu_0\approx2.91\times10^{-6}n_e\ln\Lambda T_e^{-3/2}$ and $\nu_{ei}\approx Z\nu_0$ are the respective collision rates. The effective electron temperature is defined as $T_e=(2T_\perp+T_\parallel)/3$.

The normalized collision rate $\nu_e/\omega_p\propto\sqrt{n_e}$ depends on plasma density, resulting in a decrease of the normalized collisional growth rate with density. The dashed line in Fig. \ref{fig-weibel-cp-g}(b) represents the best fit to the experimental data using this equation, giving a collisionless growth rate $\gamma_0\approx(1.5\pm0.3)\times10^{-3}\omega_p$ and $T_e\approx230\pm50$ eV. Plasma parameters, including $T_{\rm hot}\approx260$ eV, $T_{\rm cold}\approx180$ eV, and $A\approx0.48$ were retrieved using these values. The most unstable (collisionless) wave number calculated is $k\approx0.38\pm0.04\omega_p/c$, which agrees with the measured $k=0.18\omega_p/c$ within a factor of two, but may be due to collisions. The damping rate of the Weibel magnetic fields shown in Fig. \ref{fig-weibel-cp-g}(c) also exhibits a dependence on plasma density, but is still on the order of $10^{-4}\omega_p^{-1}$, indicating that the periodic magnetic fields formed upon the saturation of the Weibel instability can persist for tens of picoseconds. This low damping rate also suggests that a magnetostatic mode generated by other methods, such as by colliding a relativistic ionization front with a second EM wave \cite{lampe_interaction_1978,mori_generation_1991,wu_efficient_2023}, may endure for a sufficiently long period, rendering it useful as an ultracompact undulator \cite{fiuza_high-brilliance_2010}.

\begin{figure}[h]%
\centering
\includegraphics[width=0.5\textwidth]{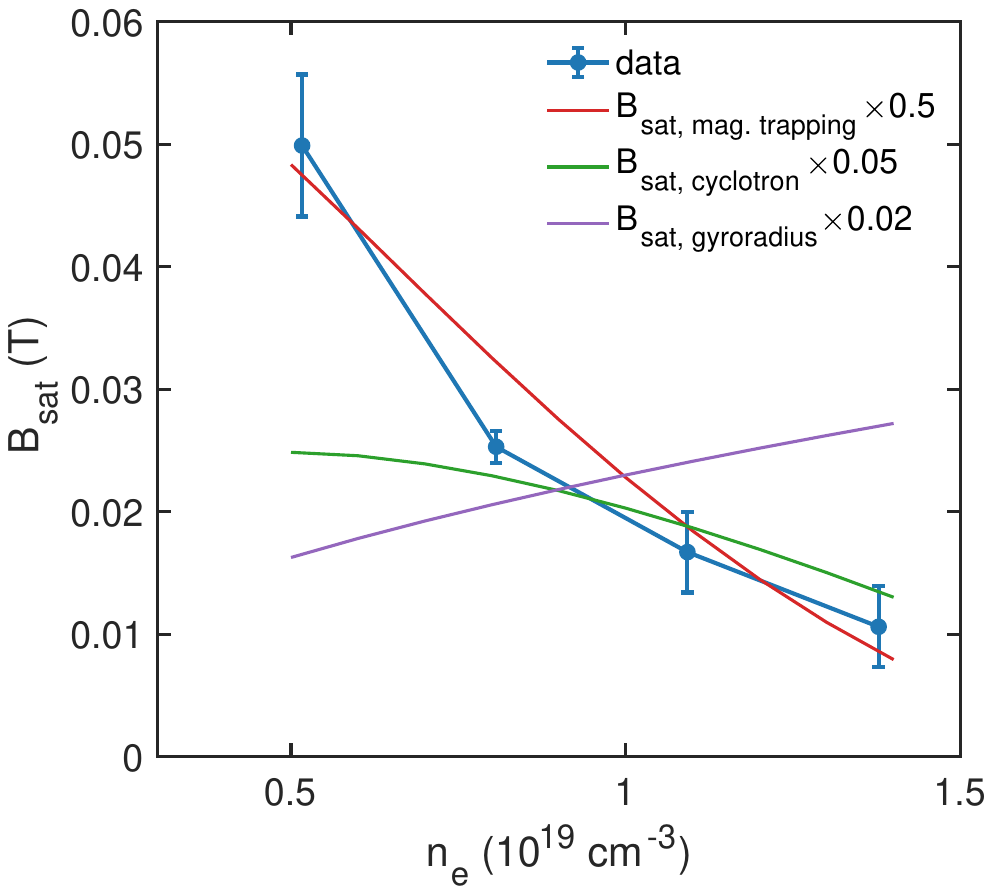}
\caption[Comparison of saturation mechanisms of Weibel instability with experimental measurements.]{Comparison of saturation mechanisms of Weibel instability with experimental measurements. The blue dots show the dependence of the measured saturated magnetic fields on plasma density. The red line represents the saturation level dependence on plasma density via the magnetic trapping mechanism, which shows the best agreement with the measurement. The green and purple lines correspond to the saturation level dependence on plasma density via equating cyclotron frequency to the growth rate and gyroradius to half of the magnetic wavelength, respectively. These mechanisms show significant deviation from the experimental data.}
\label{fig-weibel-cp-saturation}
\end{figure}

There are several mechanisms that cause the WI to saturate, one of which is magnetic trapping \cite{bret_multidimensional_2010}. This occurs when the electron bouncing frequency in the magnetic field is on the same order of magnitude as the growth rate of the instability. Based on this assumption, the saturated level of the magnetic field can be expressed as

\begin{equation}
\frac{eB_{\rm sat}}{m_ec\omega_p}\sim\left< \frac{c}{v_{\rm hot}}\right>\frac{\omega_p}{kc}\left[\frac{\gamma(k)}{\omega_p}\right]^2
\label{eqn-magnetic-trapping}
\end{equation}
where $\left<c/v_{\rm hot}\right>$ denotes the average over the particle distribution. We calculated the saturated magnetic field magnitude for different plasma densities using the measured wave number $k=0.18\omega_p/c$, growth rate $\gamma_c$, and $v_{\rm hot}\approx 0.02c$, and plotted the results in Fig. \ref{fig-weibel-cp-saturation}. To match the measurement, the calculated $B_{\rm sat}$ has been multiplied by a factor of 0.5, which indicates that the Weibel magnetic field stops growing when the electron bouncing frequency reaches about 70\% of the growth rate. There are alternative saturation mechanisms discussed in the literature. For instance, by equating the cyclotron frequency $\omega_{ce}=eB/m_ec$ to the growth rate, the saturated magnetic field can be derived as
\begin{equation}
\frac{eB_{\rm sat}}{m_ec\omega_p}\sim\frac{\gamma}{\omega_p}
\label{eqn-cyclotron}
\end{equation}

Another estimate of the saturated magnetic field is obtained by equating the gyroradius to half of the wavelength of the magnetic field, which yields the following
\begin{equation}
\frac{eB_{\rm sat}}{m_ec\omega_p}\sim\frac{kc}{\pi\omega_p}\frac{v_{\rm hot}}{c}
\label{eqn-gyro}
\end{equation}

These estimates are also plotted in Fig. \ref{fig-weibel-cp-saturation}. However, they are orders of magnitude larger and/or give the wrong trend. In other words, the measurements suggest that magnetic trapping is the dominating saturation mechanism in the photoionized plasmas used in this experiment.

\subsection{Visualization of the self-organized microscopic plasma currents}\label{sec5-visualization}
In Sec. \ref{sec3}, we elucidated that the growth of the WI in anisotropic temperature plasmas stems from the self-organization of microscopic plasma currents, where the continuous merging of the currents amplifies the magnetic fields. While prior studies have employed proton radiography to observe filamentary field structures using the colliding plasma platform (see Fig. \ref{fig-probing}), direct visualization of the microscopic plasma currents has been lacking. By employing photoionized plasma and ultrafast probing using electron bunches, we have provided a means to explore this new regime.

To visualize self-organized plasma currents, two requirements must be met. Firstly, magnetic fields with appropriate topology should be prepared to minimize integration averaging effects. This can be done by preparing a photoionized plasma with an electron velocity distribution like that in Fig. \ref{fig-an-anisotropicEVD}(a). Secondly, spatial and temporal resolution should be improved to visualize smaller features, which can be achieved by using a shorter probe bunch and electron beam optics that magnify the image. These enhancements have allowed us to visualize the self-organization process due to electron Weibel instability in photoionized plasmas with unprecedented resolutions, and validate kinetic theory predictions that may have important application to the concept of a turbulent dynamo \cite{kulsrud_critical_1999,zweibel_seeds_2013} that generates astrophysical magnetic fields in galaxies \cite{zweibel_magnetic_1997,widrow_origin_2002,kulsrud_origin_2008}.

\begin{figure}[h]%
\centering
\includegraphics[width=0.95\textwidth]{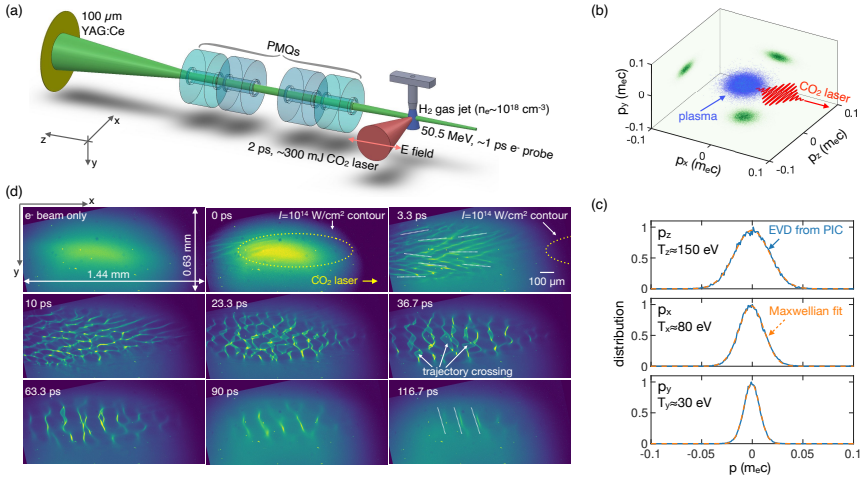}
\caption[Experimental setup, simulated initial EVD of the photoionized plasma and representative snapshots of self-generated Weibel magnetic fields.]{Experimental setup, simulated initial EVD of the photoionized plasma and representative snapshots of self-generated Weibel magnetic fields. (a) Schematic of the experimental setup. (b) Simulated distribution of plasma electrons in the 3D momentum space. (c) Projected EVDs (blue) and corresponding Maxwellian fits (red). (d) Representative frames of the electron beam deflection by self-generated magnetic fields in the plasma. The first frame shows the electron beam profile with no laser. The following frames show the evolution of the Weibel magnetic fields in the plasma. The yellow dotted ellipse on the 0 ps frame outlines the estimated $10^{14}~\rm W/cm^2$ ionization threshold intensity contour of the $\rm CO_2$ laser. The dotted white lines on the 3.3 ps and 116.7 ps frames highlight the orientation of selected density strips. On the 36.7 ps frame, the white arrows indicate structures caused by the trajectory crossing of the probe electrons.}
\label{fig-visualization}
\end{figure}

The experiment was conducted at the Accelerator Test Facility of Brookhaven National Laboratory (ATF-BNL) using a setup illustrated in Fig. \ref{fig-visualization}. Anisotropic underdense plasmas with a density of $n_e\approx(1.8\pm0.2)\times10^{18}~\rm cm^{-3}$ were created by ionizing a supersonic hydrogen gas jet using sub-terawatt, 2-ps, linearly polarized (in $z$) $\rm CO_2$ laser pulses (propagating in $x$) with a peak intensity of $1.8\times10^{15}~\rm W/cm^2$. The magnetic fields in the plasma and their spatiotemporal evolution were probed using ultrashort relativistic electron bunches delivered by the ATF linear accelerator. We used a set of permanent magnet quadruples (PMQs) to enlarge the electron image formed after the probe had traversed the plasma and relay it to the scintillator screen. The modulated electron flux was then converted to an optical image. By moving the PMQs, the image plane could be placed very close to the plasma, enabling us to improve the spatial resolving power to $\sim2.9~\rm\mu m$.

We performed a self-consistent PIC simulation to evaluate the initial EVD of the photoionized hydrogen plasma created by the CO2 laser. The simulation results are presented in Fig. \ref{fig-visualization}(b) and (c). Figure \ref{fig-visualization}(b) shows the 3D momentum space of the photoionized electrons immediately after the laser has passed. The projected distributions on different coordinates are plotted in Fig. \ref{fig-visualization}(c), which can be well-fitted by Maxwellian distributions in all three directions. Thus, we determined the plasma temperatures to be $T_z\approx150$ eV, $T_y\approx30$ eV, and $T_x\approx80$ eV. As expected from the ionization physics, the temperature along the laser polarization ($z$) direction is the highest. The temperature in the $y$ direction is the lowest, while the temperature in the laser propagation direction is in between. The increased $T_x$ compared to the temperature solely due to ionization, i.e., $T_y$, suggests that the plasma has been preferentially heated up in the longitudinal direction, possibly due to stimulated Raman scattering \cite{umstadter_observation_1987,darrow_strongly_1992,silva_anisotropic_2020}. The temperature anisotropies are $A_{zy}\equiv T_z/T_y-1\approx4$ and $A_{zx}\equiv T_z/T_x-1\approx 0.9$. These values were used to calculate the growth rate of the instability and estimate the thermal energy density of the plasma.

A movie capturing the density bunching of the electron probe beam resulting from the magnetic field deflections in the plasma was obtained by varying the delay of the electron probe relative to the $\rm CO_2$ laser. The raw data showing the density modulations on the electron beam at selected times with respect to plasma formation is presented in Fig. \ref{fig-visualization}(d). The images were acquired using PMQs with an object plane situated $10\pm0.5$ mm mm downstream of the plasma. The time origin was established at the point when the observable structures within the electron beam reached the midpoint of the field of view. The subsequent frame (captured 3.3 ps later) shows the front of the density structure has shifted to the right by approximately 1 mm, as anticipated. The total duration of the sequence covered approximately 150 ps.

The probing geometry limits the probe’s ability to sample the $B_z$ component, instead deflecting the probe via $\bf v \times B$ force along $B_x$ and $B_y$. The 0 and 3.3 ps frames in Fig. \ref{fig-visualization}(d) show the most prominent features as horizontal density strips parallel to the laser propagation direction, indicated by dotted white lines. These strips arise due to the probe electrons being deflected in the vertical $y$ direction, indicating that the dominant magnetic field component is $B_x$ with wavevector along $y$. 

In the early stage of instability, the magnetic fields evolve rapidly changing amplitude, orientation, and typical scale-length. For instance, in the 10-ps frame, the horizontal density strips in the right-hand side (front) of the frame break up into smaller-scale fish-net structures in the $x{\text-}y$ plane within 1 ps (300 microns). Fish-net structures indicate electrons are bent in both $x$ and $y$ directions, with $B_x$ and $B_y$ of approximately equal magnitude. These structures last for approximately 20 ps. As the magnetic fields continue to grow, stronger deflection of the probe electrons causes them to come to a focus before reaching the object plane of the PMQs, forming caustics \cite{levy_development_2015,patsyk_observation_2020} in the 36.7 ps frame. The density strips begin to line up in the vertical direction in this and following frames, indicating that electrons are predominantly deflected along $x$ by $B_y$. As the instability evolves, the spacing between strips increases and the structure becomes a quasi-single mode. In the final frame (116.7 ps), the field has evolved to a quasi-single mode with $\sim145~\rm\mu m$ wavelength (see dotted white lines). The magnitude of the electron probe density modulation also evolves with time, correlating with magnetic field amplitude.

\begin{figure}[h]%
\centering
\includegraphics[width=1.0\textwidth]{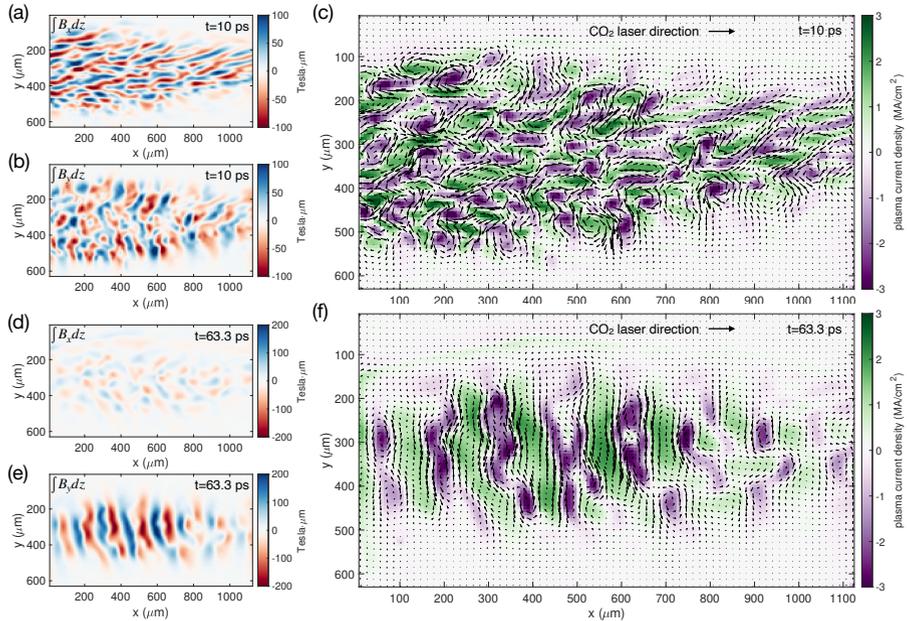}
\caption[Self-organization of microscopic plasma currents.]{Self-organization of microscopic plasma currents. (a) and (b) show the $z$-integrated $B_x$ and $B_y$ fields, respectively, for the 10 ps frame. The corresponding current density distribution, calculated from these fields, is shown in panel (c) using the color scale, while the vector magnetic field ${\bf B}_\perp={\hat x}B_x+{\hat y}B_y$ is indicated by arrows. The same quantities for the 63.3 ps frame are presented in panels (d)-(f).}
\label{fig-currents}
\end{figure}

The path integrals of magnetic fields along the probe propagation direction were retrieved by solving an equivalent optimal transport problem. Figure \ref{fig-currents}(a) and (b) show the retrieved $\int B_x{\rm d}z$ and $\int B_y{\rm d}z$ fields, respectively, for the 10-ps frame. The calculated path-integrated magnetic fields reach a peak magnitude of $\sim$100 $\rm Tesla\cdot\mu m$. The retrieved path-integrated $B_x$ and $B_y$ fields display similar peak magnitudes of $\sim0.35$ Tesla. Since the plasma is hottest along the electron probing direction, the current density $J_z$ is expected to be the dominating source for the observed magnetic fields, with the contribution of the displacement current being small. Therefore, we calculated the current density $J_z$ using the retrieved magnetic fields via $J_z=\mu_0^{-1}(\partial_xB_y-\partial_yB_x)$. The 2D distribution of the current density $J_z$ is presented in Fig. \ref{fig-currents}(c) via the color scale. The vector map of the magnetic field is overlaid onto the current density plot as arrows. As anticipated from the measured magnetic field structure, the plasma current density is modulated along both $x$ and $y$ directions, supporting the multidimensional nature of the instability.

As the instability grows, the plasma currents self-organize and merge into larger filaments. Figure \ref{fig-currents}(d)-(f) present the retrieved magnetic fields $B_x$, $B_y$, and current density $J_z$ at 63.3 ps. Now the $B_y$ field dominates over $B_x$ due to the larger initial anisotropy $A_{zy}$. The currents exhibit a more regularly spaced structure, and the current density modulation magnitude is estimated to be $\sim5\%$. In Weibel’s theory, this current density modulation is caused by the redistribution of microscopic plasma currents and does not necessarily require plasma density modulation, although density modulation may develop in the nonlinear stage of the instability. We note that in a recent experiment, Thomson scattering measurement suggested a current density modulation approaching unity in ion current filaments \cite{swadling_measurement_2020}.

\begin{figure}[h]%
\centering
\includegraphics[width=1.0\textwidth]{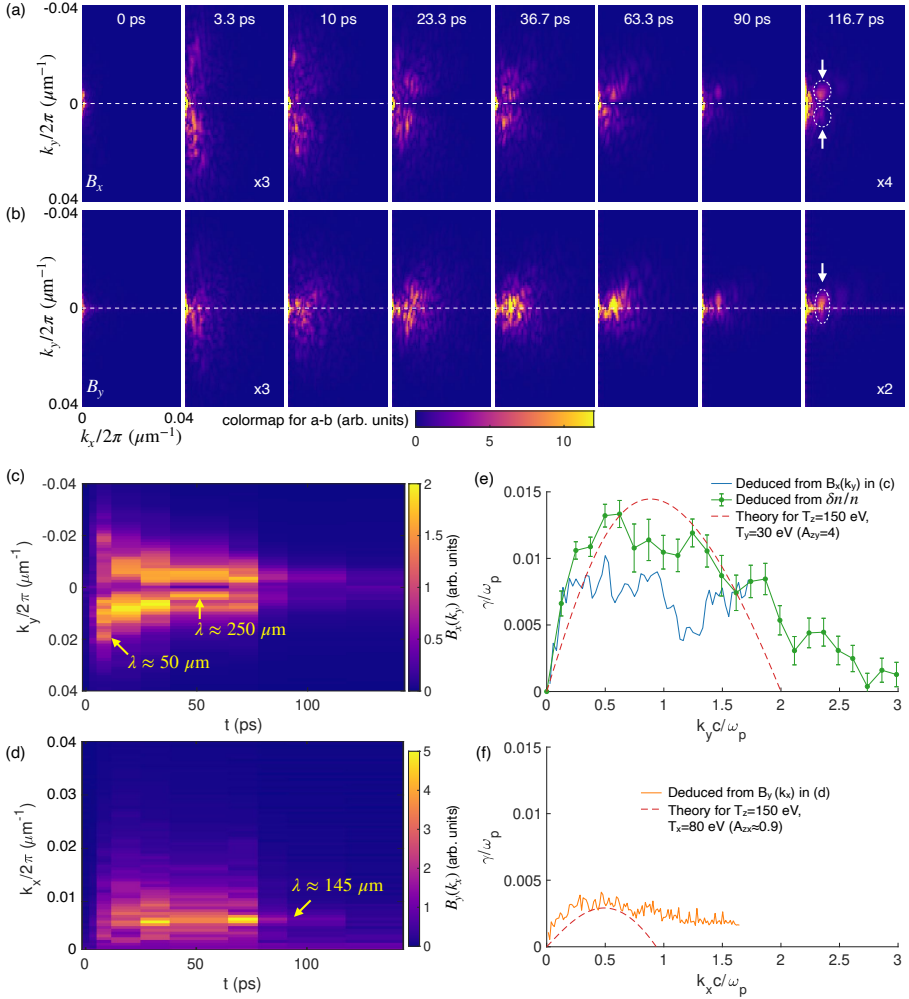}
\caption[Evolution of the $k$ spectrum of retrieved magnetic fields from experimental data.]{Evolution of the $k$ spectrum of retrieved magnetic fields from experimental data. (a) and (b) 2D $k$ spectrum for $B_x$ and $B_y$, respectively, with white circles marking surviving quasi-single mode in the last frames. (c) and (d) Temporal evolution of $k_y$ component of $B_x$ and $k_x$ component of By, respectively, with yellow arrow marking dominant mode. (e) and (f) $k$-resolved growth rates and comparison with 1D kinetic theory, with blue/green curve deduced from data in (c) and probe density modulation, and orange curve deduced from data in (d). Dashed lines in (e) and (f) show kinetic theory predictions.}
\label{fig-kspace}
\end{figure}

The process of self-organization of plasma currents was further analyzed in $k$-space by performing a 2D Fourier transform of the magnetic field components. The results of this analysis are presented in Fig. \ref{fig-kspace}. These results provide insight into the evolution of unstable modes and the transition to the dominant wavevector in 2D $k$-space. Specifically, it is observed that both $B_x$ and $B_y$ components initially exhibit broad spectra, but subsequently the unstable region continuously shrinks in size, leading to the appearance of narrow peaks that correspond to quasi-single mode formation (indicated by white circles and arrows).

The time evolution of the magnetic field spectra in the $k$-space is depicted in Fig. \ref{fig-kspace} (c) and (d) for $B_x$ and $B_y$ components, respectively. The $B_x$ field exhibits a spectral peak at $\lambda_y\approx50~\rm\mu m$ immediately after the onset of instability (at 3.3 ps), which continuously shifts towards smaller $k_y$ values, or increasing wavelength. At $\sim50$ ps, the wavelength of $B_x$ has increased to $\sim$250 $\rm\mu m$ and then remained almost constant. We speculate that this behavior is a consequence of the plasma being confined transversely, which sets an upper limit for the wavelength. The $B_y$ field initially displays a broad spectrum but converges to a wavelength of $\sim$145 $\rm\mu m$ at $\sim$30 ps, remaining nearly constant for up to $\sim$100 ps. It is worth noting that in the experiment, the wavelength of $B_y$ continued to increase with time, reaching $\sim$300 $\rm\mu m$ at $\sim$0.5 ns.

The $k$- and time-resolved data presented in Fig. \ref{fig-kspace}(c) and (d) allow us to deduce the $k$-resolved growth rates of the two magnetic field components. Specifically, each row in Fig. \ref{fig-kspace}(c) corresponds to the temporal evolution of a specific $k_y$ component of the measured $B_x$ field, and the $k$-resolved growth rate can be obtained by assuming exponential growth and fitting the data. The resulting growth rate is shown by the blue curve in Fig. \ref{fig-kspace}(e), which peaks at $k_x\approx0.5\omega_p/c$. Alternatively, the growth rate can be estimated from a single frame, such as the 3.3 ps frame in Fig. \ref{fig-visualization}(d), by tracking the increase in the density modulation magnitude of each column from right to left, which corresponds to an increasing delay. The resulting growth rate is shown by the green curve in Fig. \ref{fig-kspace}(e), and the two methods show qualitative agreement, with the intraframe method giving a slightly larger growth rate due to its higher temporal resolution. A similar analysis is applied to the $B_y$ field, and the result is shown by the orange curve in Fig. \ref{fig-kspace}(f).

Using the tri-Maxwellian $T_e(x,y,z)$ EVD obtained from PIC simulation [see Fig. \ref{fig-visualization}(b)], we can calculate the growth rate $\gamma(k)$ for the two perpendicular magnetic field components with the assumption that there is no coupling between them during the linear phase. Specifically, we utilize Eqn. \ref{eqn-disper-weibel} to perform a 1D theoretical calculation for each component. The results of these calculations are shown by the red dashed lines in Fig. \ref{fig-kspace} (e) and (f). We observe that the calculated growth rates are in reasonable agreement with the experimentally deduced growth rates for both field components, further supporting the argument that these fields are associated with the WI.

\begin{figure}[h]%
\centering
\includegraphics[width=0.95\textwidth]{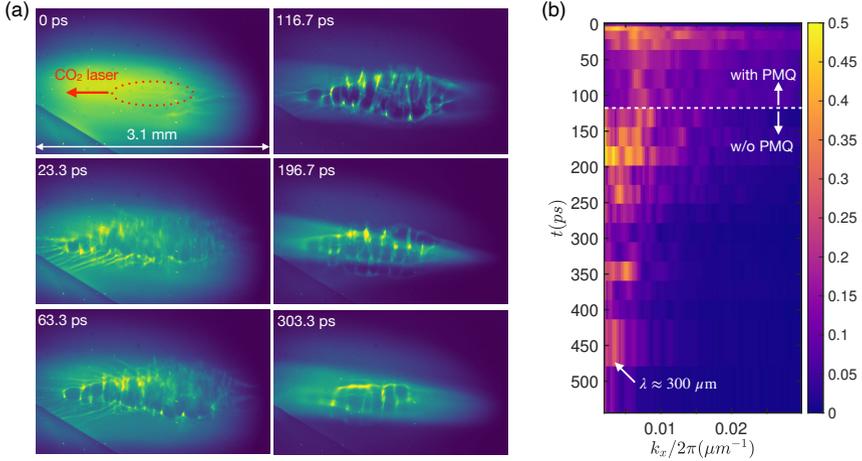}
\caption[Quasi-single mode over 0.5 ns.]{Quasi-single mode over 0.5 ns. (a) Snapshots of magnetic fields without PMQs, with the red dotted ellipse on the 0 ps frame indicating the $\rm CO_2$ laser's $10^{14}~\rm W/cm^2$ ionization threshold contour. (b) Evolution of $k$ spectra. Data for $t<117$ ps is replaced with PMQ-acquired data.}
\label{fig-half-ns}
\end{figure}

We can extend the sensitivity range of the wavenumbers $k_x$ that we can measure to even smaller values (longer wavelength) by removing the imaging PMQs to allow the probe electrons to freely propagate in vacuum after being deflected by the magnetic fields in the plasma. The probe electrons then formed images directly on the same scintillator detector. In this configuration, the magnification of the probe beam was solely due to the divergence of the beam. The main differences between the two configurations (with or without PMQs) were the object plane position and the spatial resolution. The imaging system with PMQs was suitable for measuring stronger fields and resolving shorter wavelengths, while the imaging system without PMQs could capture weaker but longer-wavelength (smaller $k_x$) fields.

We present an example dataset taken without PMQs and under the same laser and plasma parameters as in Fig. \ref{fig-visualization}(d), with time-resolved snapshots shown in Fig. \ref{fig-half-ns}(a). Notably, the enhanced sensitivity and expanded field of view allowed for the observation of additional details in the laser-plasma interaction, including filamentary structures on the left side of the field of view that were clearly visible in the 23 and 63 ps frames. However, blurry structures with shorter wavelengths appeared in the central region, which could only be resolved with imaging PMQs. Removal of the PMQs led to severe trajectory crossing before the probe electrons reached the detector, causing a washed-out region. Later, more regular structures, roughly vertical density strips, emerged and persisted for hundreds of picoseconds. Fig. \ref{fig-half-ns}(b) displays the corresponding $k$ spectrum of the density modulation as a function of time, wherein $k$ spectra for $t<117$ ps were replaced with those obtained with PMQs to compensate for considerable trajectory crossing. The stitched $k$ spectrum evolution unequivocally showed that the wavelength of the magnetic field continued to elongate to around 300 $\rm\mu m$ at half a nanosecond after plasma production. This second part was only discernible by the PMQ-out configuration, thanks to its superior sensitivity. Estimated sensitivity of PMQ-in and PMQ-out configurations were $\sim(10^3-10^1)$ and $\sim(10^1-10^{-1})$ $\rm Tesla\cdot\mu m$, respectively. The combination of these two configurations provided a dynamic range of four orders of magnitude, with a measured spatial resolution better than 3 $\rm\mu m$ (PMQ-in configuration). For instance, magnetic field measurement at 300 ps was $\sim$0.006 T, approximately two orders of magnitude smaller than that at 30 ps. Thus, our system is well-suited for exploring many scenarios involving magnetic fields in plasmas.

This temporal evolution of the range of unstable $k_x$, from an initially broad spectrum to finally a far narrower $k_x$ mode shown in Fig. \ref{fig-half-ns}(b) has not been seen before in an experiment and is consistent with picturing self-organization of the thermal Weibel magnetic fields in photoionized plasmas as a process \cite{kugland_self-organized_2012} where energy is transferred from smaller to larger scale structures.

We can estimate the fraction of kinetic energy converted to magnetic energy in our experiment. The retrieved magnetic fields have a peak rms magnitude of approximately 0.35 T, which represents a lower limit due to the probe electron trajectory crossing and possible cancellation of opposite magnetic fields along the probe direction [see the 36.7 ps frame in Fig. \ref{fig-visualization}(d)]. Analysis presented in \cite{zhang_mapping_2022} shows that the plasma beta, defined as the ratio of thermal pressure ($p_{\rm th}=n_ek_BT_e$) to magnetic pressure ($p_{\rm mag}=B^2/2\mu_0$), is approximately 100. Therefore, upon saturation, about 1\% of the thermal energy in the plasma has been converted to magnetic field energy, in agreement with our simulation results presented in \cite{zhang_mapping_2022}. This conversion rate is also consistent with previous 3D PIC simulations with quasi-relativistic temperatures \cite{romanov_self-organization_2004}, as well as PIC simulations of expanding plasmas \cite{schoeffler_magnetic-field_2014} and anisotropic plasmas driven by shearing flows \cite{zhou_spontaneous_2022}. Our findings therefore suggest that the Weibel instability may be an effective mechanism for seeding the galactic dynamo \cite{ryu_turbulence_2008}.

\section{Summary}\label{sec6}
We have presented experimental results using a novel platform for investigating the self-organization of photoionized plasmas driven by kinetic instabilities. By creating plasmas with highly non-thermal and anisotropic electron velocity distributions, we observed the emergence of a hierarchy of kinetic instabilities in plasmas ionized by circularly polarized laser pulses. Two-stream and current filamentation instabilities, driven by counter-propagating streams in the stationary plasma, were the first to appear, growing rapidly on a sub-picosecond timescale and leading to the near thermalization and isotropization of the plasma. We employed Thomson scattering measurements with femtosecond temporal resolution to verify the ultrafast growth of these instabilities, validating the kinetic theory in this previously inaccessible regime. We also observed that following the current filamentation instability, the thermal Weibel instability further amplifies the magnetic fields, resulting in magnetic fields with a macroscopic helicoid structure. We utilized electron probing with $\rm\mu m$, ps spatiotemporal resolution to measure these fields and found that they formed long lasting ($\sim$100 ps in underdense plasmas with densities on the order of $10^{18}{\text -}10^{19}~\rm cm^{-3}$) zero frequency structures.

In plasmas ionized by linearly polarized $\rm CO_2$ laser pulses, thermal Weibel instability is seen to amplify the magnetic fields from noise, resulting in quasi-static magnetic fields with different topology. Our measurements validated several predictions of kinetic theory, including simultaneous excitation of a broad wavevector spectrum, narrowing of the spectrum as the instability grows, wavenumber dependent growth rate, effects of collisions on the growth rates, topology dependence of the magnetic fields upon saturation, and saturation mechanisms. Our high spatiotemporal resolution measurements represent the first direct visualization of the self-organization of microscopic plasma currents and conclusively demonstrate the growth of thermal Weibel instability in photoionized plasmas. We showed that thermal Weibel instability is capable of amplifying magnetic fields to Tesla level and converting $\sim1\%$ of the thermal energy of the plasma into magnetic energy, suggesting that this instability may be a candidate for seeding the subsequent turbulent dynamo that is thought to be the source of observed micro-gauss level magnetic fields in galaxies.

In conclusion we have shown that details of kinetic instabilities and the resultant self-organization can be studied using high-field photoionized plasmas. We have deployed several novel diagnostic techniques to create and diagnose the needed electron velocity distribution functions and then follow the evolution of the subsequent kinetic instabilities that follow and give rise to self-organization in such plasmas. We believe that the experimental platform we have demonstrated holds immense potential for investigating coherent structures in plasmas on ultrafast timescales. In the future it will find applications in studies of magnetic field dynamics in relativistic plasmas, as well as magnetic reconnection. These topics are highly relevant to plasma astrophysics, and our platform can provide insights into the underlying physics of these phenomena in controlled experiments that can be conducted in a modest-scale laboratory.

\section*{Acknowledgement}
We acknowledge the support of the U.S. Department of Energy through Grant Nos. DE-SC0010064 and DE-SC0014043, as well as a sub-contract from Stony Brook University 72115/1126474 and the National Science Foundation through Grant No. 1734315 and 20003354. The simulations were carried out using the NERSC Cori cluster at Lawrence Berkeley National Laboratory (LBNL) under Contract No. DE- AC02-5CH11231. The authors would like to thank Dr. Ken Marsh for his help with the experiments and Prof. Warren B. Mori for many useful discussions. The authors also acknowledge the help of the ATF staff and that of Dr. Irina Petrushina and Prof. Navid Vafaei-Najafabadi of Stony Brook University.

\section*{Conflict of interest statement}
On behalf of all authors, the corresponding author states that there is no conflict of interest.


\end{document}